\documentclass[prd,aps,floats,twocolumn,nofootinbib]{revtex4}
\usepackage[usenames,dvipsnames]{color}
\usepackage{graphicx, array}
\usepackage{graphics,epsfig}
\usepackage{amssymb}
\usepackage{amsmath}
\usepackage{ulem}
\usepackage{multirow}
\usepackage{subfigure}
\usepackage{hyperref}
\usepackage{ulem}
\usepackage{bm}
\pdfoutput=1
\newcolumntype{C}[1]{>{\centering\arraybackslash}m{#1}}
\newcolumntype{N}{@{}m{0pt}@{}}
\def\be{\begin{equation}}
\def\ee{\end{equation}}
\def\bi{\begin{itemize}}
	\def\ei{\end{itemize}}
\def\ben{\begin{enumerate}}
	\def\een{\end{enumerate}}
\def\bt{\begin{tabular}}
	\def\et{\end{tabular}}
\def\bc{\begin{center}}
	\def\ec{\end{center}}
\def\cs{DR4 ILC~} 
\def\evepaper{V21} 
\def\caof{DR5 f150~} 
\def\canbare{DR5 f090} 
\def\csbare{DR4 ILC} 
\def\caofbare{DR5 f150} 
\def\can{DR5 f090~} 
\def\bea{\begin{eqnarray}}
\def\eea{\end{eqnarray}}

\def\ba{\begin{eqnarray}}
\def\ea{\end{eqnarray}}
\def\phat{\hat{p}}

\makeatletter
\def\p@subsection{}
\makeatother
\begin{document}
\input{epsf}
\title{The Atacama Cosmology Telescope: Detection of the Pairwise Kinematic Sunyaev-Zel'dovich Effect with SDSS DR15 Galaxies}

\author{V.~Calafut$^1$}
\author{P.~A.~Gallardo$^2$}
\author{E.~M.~Vavagiakis$^2$}
\author{S.~Amodeo$^1$}
\author{S.~Aiola$^{3}$}
\author{J.~E.~Austermann$^4$}
\author{N.~Battaglia$^1$}
\author{E.~S.~Battistelli$^5$}
\author{J.~A.~Beall$^4$}
\author{R.~Bean$^1$}
\author{J.~R.~Bond$^6$}
\author{E.~Calabrese$^7$}
\author{S.~K.~Choi$^{2,1}$}
\author{N.~F.~Cothard$^8$}
\author{M.~J.~Devlin$^9$}
\author{C.~J.~Duell$^2$}
\author{S.~M.~Duff$^4$}
\author{A.~J.~Duivenvoorden$^{10}$}
\author{J.~Dunkley$^{10,11}$}
\author{R.~Dunner$^{12}$}
\author{S.~Ferraro$^{13,14}$}
\author{Y.~Guan$^{15}$}
\author{J.~C.~Hill$^{3,16}$}
\author{G.~C.~Hilton$^4$}
\author{M.~Hilton$^{17}$}
\author{R. Hlo\v{z}ek$^{18,19}$}
\author{Z.~B.~Huber$^2$}
\author{J.~Hubmayr$^4$}
\author{K.~M.~Huffenberger$^{20}$}
\author{J.~P.~Hughes$^{21}$}
\author{B.~J.~Koopman$^{22}$}
\author{A.~Kosowsky$^{15}$}
\author{Y.~Li$^2$}
\author{M.~Lokken$^{6,18,19}$}
\author{M.~Madhavacheril$^{23}$}
\author{J.~McMahon$^{24,25,26,27}$}
\author{K.~Moodley$^{17,28}$}
\author{S.~Naess$^{3}$}
\author{F.~Nati$^{29}$}
\author{L.~B.~Newburgh$^{22}$}
\author{M.~D.~Niemack$^{2,1}$}
\author{L.~A.~Page$^{10}$}
\author{B.~Partridge$^{30}$}
\author{E.~Schaan$^{12,13}$}
\author{A.~Schillaci$^{31}$}
\author{C.~Sif\'on$^{32}$}
\author{D.~N.~Spergel$^{3,11}$}
\author{S.~T.~Staggs$^{10}$}
\author{J.~N.~Ullom$^4$}
\author{L.~R.~Vale$^4$}
\author{A.~Van~Engelen$^{33}$}
\author{J.~Van Lanen$^4$}
\author{E.~J.~Wollack$^{34}$}
\author{Z.~Xu$^{9,35}$}

\affiliation{$^1$ Department of Astronomy, Cornell University, Ithaca, NY 14853, USA}
\affiliation{$^2$ Department of Physics, Cornell University, Ithaca, NY 14853, USA}
\affiliation{$^{3}$ Center for Computational Astrophysics, Flatiron Institute, New York, NY 10010, USA }
\affiliation{$^4$ NIST Quantum Devices Group, 325 Broadway Mailcode 817.03, Boulder, CO 80305, USA }
\affiliation{$^5$ Physics Department, Sapienza University of Rome, Piazzale Aldo Moro 5, I-00185, Rome, Italy}
\affiliation{$^6$ Canadian Institute for Theoretical Astrophysics, University of Toronto, Toronto, ON, M5S 3H8, Canada}
\affiliation{$^7$ School of Physics and Astronomy, Cardiff University, The Parade, Cardiff, CF24 3AA, UK}
\affiliation{$^8$ Department of Applied and Engineering Physics, Cornell
University, Ithaca, NY, USA 14853}
\affiliation{$^{9}$ Department of Physics and Astronomy, University of Pennsylvania, 209 South 33rd Street, Philadelphia, PA 19104, USA }
\affiliation{$^{10}$ Joseph Henry Laboratories of Physics, Jadwin Hall, Princeton University, Princeton, NJ, USA 08544}
\affiliation{$^{11}$ Department of Astrophysical Sciences, Peyton Hall, Princeton University, Princeton, NJ USA 08544}
\affiliation{$^{12}$ Instituto de Astrof\'isica and Centro de Astro-Ingenier\'ia, Facultad de F\'isica, Pontificia Universidad Cat\'olica de Chile, Av. Vicu\~na Mackenna 4860, 7820436, Macul, Santiago, Chile}
\affiliation{$^{13}$ Lawrence Berkeley National Laboratory, One Cyclotron Road, Berkeley, CA 94720, USA}
\affiliation{$^{14}$ Berkeley Center for Cosmological Physics, UC Berkeley, CA 94720, USA}
\affiliation{$^{15}$ Department of Physics and Astronomy, University of Pittsburgh, Pittsburgh, PA 15260, USA }
\affiliation{$^{16}$ Department of Physics, Columbia University, New York, NY 10027, USA }
\affiliation{$^{17}$ Astrophysics Research Centre, University of KwaZulu-Natal, Westville Campus, Durban 4041, South Africa}
\affiliation{$^{18}$ David A. Dunlap Department of Astronomy and Astrophysics, University of Toronto, 50 St. George St., Toronto, ON M5S 3H4, Canada}
\affiliation{$^{19}$ Dunlap Institute for Astronomy and Astrophysics, University of Toronto, 50 St. George St., Toronto, ON M5S 3H4, Canada}
\affiliation{$^{20}$ Department of Physics, Florida State University, Tallahassee FL, USA 32306}
 \affiliation{$^{21}$ Department of Physics and Astronomy, Rutgers, the State University of New Jersey, 136 Frelinghuysen Road, Piscataway, NJ 08854, USA}
\affiliation{$^{22}$ Department of Physics, Yale University, New Haven, CT 06511, USA}
\affiliation{$^{23}$ Centre for the Universe, Perimeter Institute for Theoretical Physics, Waterloo, ON, N2L 2Y5, Canada}
\affiliation{$^{24}$ Department of Physics, University of Chicago, Chicago, IL 60637, USA}
\affiliation{$^{25}$ Department of Astronomy and Astrophysics, University of Chicago, 5640 S. Ellis Ave., Chicago, IL 60637, USA}
\affiliation{$^{26}$ Kavli Institute for Cosmological Physics, University of Chicago, 5640 S. Ellis Ave., Chicago, IL 60637, USA}
\affiliation{$^{27}$ Enrico Fermi Institute, University of Chicago, Chicago, IL 60637, USA}
\affiliation{$^{28}$ School of Mathematics, Statistics and Computer Science, University of KwaZulu-Natal, Westville Campus, Durban 4041, South Africa}
\affiliation{$^{29}$ Department of Physics, University of Milano-Bicocca, Piazza della Scienza 3, 20126 Milano (MI), Italy}
\affiliation{$^{30}$ Department of Physics and Astronomy, Haverford College, Haverford, PA 19041, USA }
\affiliation{$^{31}$ Department of Physics, California Institute of Technology, Pasadena, CA 91125, USA}
\affiliation{$^{32}$ Instituto de F\'isica, Pontificia Universidad Cat\'olica de Valpara\'iso, Casilla 4059, Valpara\'iso, Chile}
\affiliation{$^{33}$ School of Earth and Space Exploration, Arizona State University, Tempe, AZ 85287,  USA }
\affiliation{$^{34}$ NASA Goddard Space Flight Center, Greenbelt MD 20771}
\affiliation{$^{35}$ MIT Kavli Institute, Massachusetts Institute of Technology, Cambridge, MA 02139, USA }
\label{firstpage}


\begin{abstract}
\clearpage
We present a 5.4$\sigma$ detection of the pairwise kinematic Sunyaev-Zel’dovich (kSZ) effect using Atacama Cosmology Telescope (ACT) and {\it Planck} CMB observations  in combination with Luminous Red Galaxy samples from the Sloan Digital Sky Survey (SDSS) DR15 catalog. Results are obtained using three ACT CMB maps: co-added 150~GHz and  98~GHz maps, combining observations from 2008-2018 (ACT DR5), which overlap with SDSS DR15 over 3,700 sq. deg., and a component-separated map using  night-time only observations from 2014-2015 (ACT DR4), overlapping with SDSS DR15 over 2,089 sq. deg. Comparisons of the results from these three maps provide consistency checks in relation to potential frequency-dependent foreground contamination. A total  of 343,647 galaxies are used as tracers to identify and locate galaxy groups and clusters from which the kSZ signal is extracted using aperture photometry. We consider the impact of  various aperture photometry assumptions and covariance estimation methods on the signal extraction. Theoretical predictions of the pairwise velocities are used to obtain best-fit, mass-averaged, optical depth estimates for each of five luminosity-selected tracer samples. A comparison of the kSZ-derived optical depth measurements obtained here to those derived from the thermal SZ effect for the same sample is presented in a companion paper.
 
\end{abstract}
\maketitle
\section{Introduction}
\label{sec:intro}
Deciphering the origins of accelerated cosmic expansion \citep{Perlmutter:1998np, Riess:2004nr} is one of the central goals of modern cosmology. The effects of dark energy only manifest indirectly, through possible deviations from the predictions of General Relativity (GR) and the gravitational properties of Standard Model particles and dark matter. To determine if dark energy  is a cosmological constant, a novel type of matter, or evidence that gravity deviates from GR on cosmic scales, one is principally reliant  on three cosmological tracers of the gravitational field: the positions and velocities of massive objects and the distortion they create in the geodesic paths of light from more distant objects.  

As Cosmic Microwave Background (CMB) photons traverse through a galaxy cluster they interact with the hot cluster gas, and the peculiar motion of the cluster relative to the CMB rest-frame creates a Doppler-shift in the CMB known as the kinematic Sunyaev-Zel'dovich effect (kSZ)\cite{1980MNRAS.190..413S}. Concurrently with the kSZ, the CMB photons are also heated up by the cluster gas, the thermal Sunyaev-Zel'dovich effect (tSZ).  The tSZ imprint has a characteristic frequency dependence, and can be isolated through the use of multi-frequency measurements. By contrast, the kSZ effect is an order of magnitude smaller and has a thermal spectrum that makes its detection, and separation from tSZ and dust emission foregrounds, challenging.

On scales of the order of $\sim$25-50~Mpc, the gravitational attraction between  clusters  (and groups) of galaxies causes them, on average, to move towards each other.  This pairwise motion can be used to extract the kSZ effect.  A pairwise correlation statistic \cite{1983ApJ...267..465D} is a useful approach to extracting kSZ signals because of its dependence on  differences of measured temperatures on the sky at the positions of clusters, averaging out contaminating signals like the tSZ signal and dust emission. The pairwise kSZ momentum, sensitive to both the cluster peculiar velocity and optical depth, has been shown to have the potential to probe the large scale structure (LSS) growth rate, providing insights into the evolution of dark energy, cosmic modifications to gravity over cosmic time, and constraints on the sum of the neutrino masses \cite{DeDeo:2005yr,Bhattacharya:2006ke,Kosowsky:2009nc,Bull:2011wi,Mueller:2014dba,Mueller:2014nsa,Flender:2015btu}.

Extraction of the kSZ signal is aided by using galaxy surveys to provide bright tracer galaxies to identify and locate the clusters  \cite{Kashlinsky:2009dw, Kashlinsky:2000xk, 1538-4357-515-1-L1, 2013MNRAS.430.1617L, 2017JCAP...01..057S}.
The first measurement of the kSZ signal was made by Hand et al. \cite{2012PhRvL.109d1101H} (herein H12) by estimating the mean pairwise cluster momentum with the ACT data from  2008 to 2010 observing seasons and a sample of clusters traced by galaxies in the Sloan Digital Sky Survey Data Release 9 (SDSS DR9) galaxy catalog.  This measurement has since been improved with a $4.1\sigma$ measurement in the mass-averaged optical depth, $\bar\tau$,  using improved data from ACT DR3 and  SDSS DR11 data \cite{DeBernardis:2016pdv} (herein DB17).  Detections using this  estimator have also been reported by the {\it Planck} collaboration using galaxies from SDSS \cite{Ade:2015lza}, and the South Pole Telescope collaboration using galaxies from the Dark Energy Survey \cite{Soergel:2016mce}. 

In addition to the pairwise statistics, other complementary techniques have also been applied  to measure the kSZ effect \cite{Smith:2018bpn},  including velocity reconstruction \cite{Schaan:2015uaa}, projected fields \cite{Hill:2016dta,PhysRevD.94.123526}, cross-correlation of angular redshift fluctuations \cite{Chaves-Montero:2019isa} and cluster stacking 
\cite{Tanimura:2020une}. Two recent papers \cite{Schaan:2020qhk,Amodeo:2020mmu}  focused on using velocity reconstruction and stacking of galaxy cluster samples to study the radial profiles of tSZ and kSZ signals in the ACT data.  The work used the same coadded  \cite{Naess:2020wgi} and component-separated \cite{2020PhRvD.102b3534M} maps, as are used here, but the galaxy samples are different, with different host halo masses. As a result,  the findings from these papers are not directly comparable to those in this work, nor those in the companion paper \cite{Vavagiakis:2021ilq} (\evepaper).  We find, however, that the rough signal-to-noise ratios are comparable. Overall, these two sets of  papers provide complementary ways to analyze  tSZ and kSZ effects.

Our work is laid out as follows: In Section \ref{sec:data}, we describe the ACT and {\it Planck} CMB data  and the SDSS galaxy samples used in our analysis. In Section \ref{sec:form}, we lay out the formalism for the pairwise estimator, the covariance techniques, mass-averaged optical depth fitting and signal-to-noise estimation. In Section \ref{sec:analysis}, we present our results and discuss the pairwise kSZ detections and  mass-averaged optical depth constraints.  The findings are drawn together in the Conclusion, in Section \ref{sec:conclusions}. 
 
\section{Datasets}
\label{sec:data}

\subsection{ACT data}
\label{sec:data:cmb}

 In our analysis, we use three CMB datasets that combine ACT and {\it Planck} data.  The first dataset is a component-separated internal linear combination map   (ILC) \cite{2020PhRvD.102b3534M}, referred to as \csbare, which uses night-time ACT observations  from DR4, principally from 2014 to 2015 \cite{Choi:2020ccd,Aiola:2020azj} as well as  {\it Planck} data  in eight bands, from 30 to 545~GHz, from the PR2 (2015) release \cite{2014A&A...561A..97P}. The map is created by minimizing the variance and is dominated by CMB and kSZ signals but also has other foregrounds including thermal SZ and Cosmic Infrared Background (CIB) contributions.
  
The second and third datasets are the co-added ACT DR5 98~GHz and 150~GHz maps  \cite{Naess:2020wgi} which combine ACT observations from 2008-2018 seasons, including day-time data, and  {\it Planck} PR2 \cite{Adam:2015rua}  data release centered at 100 and 143 GHz. We refer to these two maps as \can and \caofbare, respectively, using the frequency naming conventions in Naess et al. \cite{Naess:2020wgi}. The CMB maps  have point source and galactic plane masks, and a noise threshold cut of 45 $\mu$K, relative to the CMB, as discussed in more detail in \evepaper. The companion paper also includes a map of the specific regions utilized for \csbare, \caof and \can CMB and SDSS surveys.

The use of co-added maps at two different frequencies and the multi-frequency component separated map facilitates the comparison of the extracted kSZ measurements from maps in which potential thermal SZ and other secondary foreground contributions will vary.

 \begin{table*}[t!]
  	\begin{tabular}{ | C{3.5em} | C{7.5em} | C{10em} | C{4.0em} | C{4.0em} |  C{4.0em}    |C{4.0em} | C{4.0em} | C{4.0em}|}
\cline{4-9}
  		 \multicolumn{3}{l|}{}
		 &  \multicolumn{3}{c|}{\caof and \can} &  \multicolumn{3}{c|}{\cs} 
		 \\ \hline
Bin &		Luminosity cut  &  Mass cut $M_{200}$&  \multirow{2}{*}{$N_{\mathrm{gal}}  $}  & $\langle L\rangle$ &\multirow{2}{*}{$\langle{z}\rangle$ } &  \multirow{2}{*}{$N_{\mathrm{gal}}  $}    & $\langle{L}\rangle$ &\multirow{2}{*}{$\langle{z}\rangle$ } 
		\\   
Label & 		 ($10^{10}L_{\odot}$) & $ (10^{13}M_{\odot})$  &  & ($10^{10}L_{\odot}$)  & & & ($10^{10}L_{\odot}$) & 
		  \\ \hline
L43D & $4.3<L<6.1$ 	&$0.55<M<1.00$ 	& 130,577		& 5.2  	& 0.48 	& 71,699 & 5.2 & 0.48
  \\ 
L61D & $6.1<L<7.9$  	& $1.00<M<1.66$	 & 109,911	& 6.9 	& 0.48	& 61,024 &  6.9 & 0.48
\\
L43 & $L>4.3$  		&$M>0.52$		& 343,647 	&7.4	 	& 0.49 	& 190,551 & 7.4 & 0.50
  \\ 
L61 & $L>6.1$ 			&$M>1.00$		& 213,070  	& 8.7 	& 0.51	& 118,852	 & 8.7 & 0.51
  \\  
L79 & $L>7.9$ 			& $M>1.66$  		& 103,159 	& 10.6 	& 0.53	& 57,828	& 10.9 & 0.54
  \\ \hline
  	\end{tabular}
  	\caption{Summaries of the five luminosity-determined samples analyzed  in this paper along with the bin labels with which we will refer to them throughout.  The host halo mass ranges, $M_{200}$, the number of galaxies, $N_{\mathrm{gal}}$, the mean redshift, $\langle{z}\rangle$, and mean luminosity, $\langle{L}\rangle$, are given for the samples that overlap with the \caofbare, \can and \cs maps. These galaxy selection and halo mass estimates are derived in the companion paper \evepaper. }

  	\label{tab1}
  \end{table*} 

\subsection{SDSS data}
\label{sec:data:lss}

As in the previous ACT analyses, H12 and DB17, we utilize spectroscopically-selected luminous red galaxies (LRGs) as tracers of the galaxy groups and clusters in which the kSZ is to be measured. Galaxies from  SDSS DR15 \cite{2019ApJS..240...23A} are identified in the regions overlapping with the CMB maps, for \caof and \can maps the overlap is  3,700 sq. deg. while for the \cs map the overlap is 2,089 sq. deg.   Galaxies are selected  based on multiband de-redenned model  magnitudes.  Full details of the selection process, including the SDSS query, are provided in \evepaper.  The query yields  602,461 galaxies for which the luminosities are calculated from the model magnitude using K-corrections made with the \url{k_correct}\footnote{\url{http://kcorrect.org}} software \cite{Blanton:2006kt} using the SDSS asinh magnitude conversion \cite{Lupton:1999pt}.  

Samples are selected based on luminosities, with minimum thresholds $4.3$, $6.1$ and $7.9 \times 10^{10}L_{\odot}$, with the latter two chosen to align with selection criteria in DB17.  Additional galaxy cuts are applied based on ACT CMB map noise, point source excision, and removal of the galactic plane. The final catalog analyzed includes 343,647 galaxies with $L >4.3\times 10^{10} L_{\odot}$.  

We consider the kSZ properties in five luminosity bins:   two disjoint luminosity bins, $4.3<L (10^{10}L_{\odot})<6.1$,  $6.1<L(10^{10}L_{\odot})<7.9$ (referred to as L43D and L61D respectively), two  cumulative luminosity bins, $L(10^{10}L_{\odot} >4.3$ and $>6.1$ (L43 and L61, respectively) and one bin that is both disjoint and cumulative,   $L(10^{10}L_{\odot})>7.9$ (referred to as L79). The characteristics of the samples in each  luminosity-selected bin  are summarized in  Table \ref{tab1}, including the host group/cluster mass ranges,  the mean redshift and  luminosities, and the number of galaxies included in each bin. The full redshift distributions of the galaxy samples are shown in the companion paper, \evepaper.

\section{Formalism}
\label{sec:form}
  
\subsection{Pairwise momentum estimator}
\label{sec:form:pw}
The CMB temperature shift induced by the peculiar motion of a galaxy group/cluster is given by \cite{1970Ap&SS73S},
\begin{equation} 
\frac{\delta T_{\mathrm{kSZ}}}{T_0}({\bf\hat{r}}) = -\int dl\,\sigma_T \, n_e  \frac{{\bf v}\cdot {\bf\hat r}  }{c} 
\end{equation} 
where $n_e$ is the electron number density, $T_0=2.726K$ is the average CMB temperature and $\sigma_T$ is the Thomson cross-section. A positive peculiar velocity, {\bf v}, relates to motion away from the observer, so induces a negative kSZ effect. 

The temperature is obtained through aperture photometry (AP), in which the temperatures of pixels within a disk of aperture size $\Theta$ and an annulus of  equal area, out to radius $\sqrt{2}\Theta$, are differenced around each group/cluster.   We  use the positions of the tracer LRGs to center the aperture, under the assumption of the Central Galaxy Paradigm \citep{vandenBosch:2005} that the brightest galaxy within a group/cluster traces the minimum of the gravitational potential well.  

The aperture temperature is calculated by analyzing a postage stamp region centered at the angular position of the $i^{\mathrm{th}}$ galaxy,  ${\bf r}_i=\{ {\bf\hat{r}}_i, z_i\}$ that includes, but extends beyond, the group/cluster in question.
Within the postage stamp  a finer resolution pixel grid is created, 10 times smaller than the pixel size; temperatures are assigned to the finer pixels using a Fourier domain interpolation. The average temperatures of the smaller pixels contained in the disk/ring are then used to calculate the $T_{AP}(r_i,z_i,\Theta)=\bar{T}_{\mathrm{disk}}-\bar{T}_{\mathrm{annulus}}$. We checked that this gives an equivalent result to taking weighted averages of the full size pixels when  a fractional weighting equivalent to the area of each pixel within the disk or annulus is included. 

We use an aperture size of $\Theta=2.1'$, aligned with the anticipated angular size of groups/clusters in the redshift ranges we are analyzing ($2.1'$ at $z=0.5$ relates to a comoving scale of $\sim$1.1Mpc for the cosmological model assumed in the analysis, as described in Sec.~\ref{sec:form:sn}).

We estimate the kSZ temperature by calculating the temperature decrement around each tracer galaxy,
\begin{equation}
\delta T_i (r_i, z_i, \sigma_z, \Theta) = T_{AP}(r_i,z_i,\Theta) - \bar T_{AP}(r_i, z_i, \Theta, \sigma_z),
\label{eq:delT}
\end{equation} 
where, following H12 and DB17, we subtract a redshift-smoothed aperture temperature, $\bar{T}_{AP}$, to remove potential redshift dependent contamination that could mirror a pairwise signal when differencing aperture temperatures from objects separated in redshift. A Gaussian smoothing is applied for each pair using a redshift smoothing parameter, $\sigma_z$:
 \bea
  \bar T_{AP}(r_i, z_i, \Theta, \sigma_z) &=&\frac{\sum_j T_{AP} (r_i, \Theta) \exp\left(- \frac{(z_i-z_j)^2}{2\sigma_z^2}\right)}{\sum_j \exp\left(- \frac{(z_i-z_j)^2}{2\sigma_z^2}\right)}\hspace{0.5cm}
  \label{eq:TAPbar}
 \eea
We use $\sigma_z= 0.01$, as used by the {\it Planck} team \cite{Ade:2015lza} and in DB17. We demonstrate that the pairwise results are insensitive to the precise value of $\sigma_z$ in Appendix~\ref{sec:comparison}.  

Analogous with the corrections made to the tSZ temperatures in \evepaper, we correct for the difference in kSZ aperture temperature due to the differences in beam size between  the \can map (FWHM=2.1') and the \cs map (FWHM=1.6') versus the \caof map (FWHM=1.3')\citep{Naess:2020wgi}. We consider a fiducial kSZ density profile for the average virial group/cluster in each luminosity bin and  derive estimates of the kSZ signal when convolved with the respective beams for the \caof and \can maps from \cite{Naess:2020wgi}, using \texttt{Mop-c GT} \footnote{\url{https://github.com/samodeo/Mop-c-GT}}, and using a Gaussian beam  for the \cs map  (see a  detailed description in \cite{Amodeo:2020mmu}). A resulting relative beam correction factor is applied that increases  the \can map  $T_{AP}$ measurements  by 31\%  and reduces the \cs map by 5\%. 

We implement the pairwise momentum estimator \cite{1538-4357-515-1-L1} for the correlation of the velocities, 
\begin{equation}  
\phat(r) = - \frac{\sum_{i<j} (\delta T_i - \delta T_j)c_{ij}}{\sum_{i<j}  c^2_{ij}}, 
\label{eq:phat}
\end{equation} 
where the sum is over all pairs, each separated by a distance $r=|{\bf r}_{ij}|=|{\bf r}_i-{\bf r}_j|$.
The weights $c_{ij}$ are  geometrical factors that account for the alignment of a pairs $i$ and $j$ along the line of sight \cite{DeBernardis:2016pdv}, given by
\begin{equation} 
c_{ij} =  {\bf \hat r_{ij}} \cdot \frac{{\bf \hat r_i} + {\bf \hat r_j}}{2} = \frac{(r_i - r_j)(1+\cos\alpha)}{2\sqrt{r_i^2 + r_j^2 - 2r_i r_j \cos\alpha}}
\end{equation} 
where $\alpha$ is the angle between  unit vectors $\hat r_i$ and $\hat r_j$.

We analyze data in radial separation bins of width 10~Mpc centered on $r=5$ up to 145~Mpc and then four unevenly spaced bins, centered on 175, 225, 282.5 and 355~Mpc (for which the maximum included separation is 395~Mpc). The latter bins have broader widths to  account for increased correlation between spatial scales as one goes to larger separations,  as was found in DB17, and  discussed  in Appendix \ref{sec:app:covest}.

We update the kSZ pipelines used  in \cite{Calafut:2017mzp}, which analyzed {\it Planck} SEVEM maps in HEALPix format, and  in DB17. The pipeline used in this work is publicly available \footnote{\url{https://github.com/patogallardo/iskay}} and  parallelized and distributed \cite{numba,dask} in Python, and uses \texttt{Pixell}\footnote{\url{https://github.com/simonsobs/pixell}} subroutines to analyze the CMB map.

Our  aperture photometry assumptions  include some analytic differences from those in DB17: we include fractional pixel weighting, reproject the pixelation of the submap, and implement cluster-centered, instead of pixel-centered aperture photometry. We discuss these differences,  and their respective implications for the signal extraction in Appendix~\ref{sec:comparison}. 


To estimate the covariance, at least four resampling strategies have been proposed in the literature, with error bar estimates that vary up a factor of two among them (see Appendix in \citep{Soergel:2016mce} for more detail). As shown in \citep{Soergel:2016mce} there are systematic differences in the inference method that tend to dominate the uncertainty estimation. 

In cross-correlating the maps with the galaxy sample, contributions will be picked up from the residual foregrounds that are correlated with the tracer. In kSZ estimators, which are velocity weighted,  the positive and negative contributions from cross-correlation with the velocity field are not mimicked by other residual foreground contributions and their effect is suppressed effectively contributing to noise,  and not bias, in the pairwise signal. It is important to capture this noise contribution (for example from the residual tSZ)  in the covariance calculation. Estimators that sample the maps directly will capture this more effectively than simulations. 
%
%
In this work we use bootstrap estimation from the maps directly to evaluate the covariance used in the analysis. In Appendix~\ref{sec:app:covest}, we summarize the findings of the covariance estimation comparison across three different methods, covariances of simulated maps, jackknife (JK) estimation, the primary method used in \citep{DeBernardis:2016pdv,Soergel:2016mce}, and bootstrap estimation, and motivate why we use the bootstrap derived estimates in the main analysis.

\subsection{Signal to Noise and $\bar\tau$ estimates}
\label{sec:form:sn}

A theoretical prediction for the observed pairwise momentum can be modeled in terms of a mass-averaged pairwise peculiar velocity, $V$
\begin{equation}
\phat_{\mathrm{th}}(r,z)=-\frac{ T_{CMB}}{c} \bar\tau V(r,z) ,
\label{eq:pksz}
\end{equation}
where $\bar\tau$ is an effective mass-averaged measure of the  optical depth over the group/cluster samples and $z$ is taken as the mean redshift of each luminosity sample as given in Table ~\ref{tab1}.

The theoretical pairwise velocity, $V$, can be derived in terms of the correlation function \cite{1999ApJ...518L..25J}, and is calculated here following \cite{Mueller:2014nsa,Mueller:2014dba} using  linear theory \cite{Sheth:2000ff},
\bea
V(r,z) &=& -\frac{2}{3} \frac{f(z) H(z) r }{1+z} \frac{\bar\xi_h(r,z)}{1+\xi_h(r,z)},
\label{eq:vth}
\eea
where $f(z)$ is the linear growth rate and $H(z)$ is the Hubble rate. $\xi_h$  and $\bar\xi_h$ are, respectively, the 2-point halo correlation function and volume averaged halo correlation function:
\bea
\xi_{h}(r,z)&=&\frac{1}{2 \pi^2}\int dk k^2 j_0(kr) P(k,z) b_{h}^{(2)}(k), \hspace{0.5cm}
\\
\bar{\xi}_{h}(r,z)&=&\frac{3}{ r^3} \int_0^{r} dr' r'^2 \xi(r,z) b_{h}^{(1)}(k).
\eea
Here $P(k,z)$ is the linear matter power spectrum, $j_0(x)=\sin(x)/x$ is the zeroth order spherical Bessel function, and $b^{(q)}_{h}$, are mass-averaged halo bias moments 
given by,
\bea
b^{(q)}_{h}(z)&=&\frac{\int_{M_{\mathrm{min}}}^{M_{\mathrm{max}}} dM \ M \ n(M,z)b^{q}(M)W^2[kR(M,z)]}{\int_{M_{\mathrm{min}}}^{M_{\mathrm{max}}} dM\ M \ n(M,z)W^2[kR(M,z)]} \nonumber
\\ \label{eq:bq}
\eea
with $n(M,z)$  is number density of halos of mass $M$, for which we  use a halo mass function in \cite{2011ApJ...732..122B}, and the top-hat window function is given by $W(x)=3(\sin x-x \cos x)/x^3$. $R$ is the characteristic scale of a halo of mass M,  $R(M,z) =[3M/4\pi \bar\rho(z)]^{1/3}$, with $\bar\rho$ the background cosmological matter density. The lower mass limit, $M_{\mathrm{min}}$, is  taken to be the halo mass cut given in Table~\ref{tab1}. The upper mass limit, $M_{\mathrm{max}}$,  is taken to be $10^{16}M_{\odot}$. We consider in the analysis the sensitivity of the results to these specific limits, with the understanding that bias the moments are  dominated by the, far more numerous, lower mass halos.

We compare our pairwise kSZ momentum measurements to theoretical peculiar velocity predictions using a modified version of the CAMB code \citep{Lewis:2002ah} that calculates the mass-averaged pairwise velocity, $V$,  as described  in Mueller et al.\cite{Mueller:2014nsa,Mueller:2014dba}. We assume a {\it Planck} cosmology for a flat universe \cite{Ade:2015xua}: $\Omega_bh^2 = 0.02225$, $\Omega_ch^2 = 0.1198$, $H_0$ = 67.3kms$^{-1}$Mpc$^{-1}$, $\sigma_8 = 0.83$, $n_s = 0.964$. We translate the observational galaxy luminosity cuts to group/cluster mass cuts using mass-luminosity relationship described in \evepaper.

We determine the likelihood of $\bar\tau$ using the  $\chi^2$,
\begin{eqnarray}
\chi^2(\bar\tau)&=&\sum_{ij}\Delta\phat{_i}(\bar\tau)  \hat C_{ij}^{-1} \Delta\phat{_j}(\bar\tau),
\end{eqnarray}
with
 $\Delta\phat_i(\bar\tau) = \phat_{i,\mathrm{th}}(\bar\tau)-\phat_{i,\mathrm{obs}},$
%
 where $\phat_{i,th}(\bar\tau)$ is the theoretical kSZ pairwise momentum estimate at cluster separation $r_i$ for an assumed mass-averaged optical depth, $\bar\tau$, and $\phat^{\mathrm{obs}}_i$ are the measurements obtained from the ACT and SDSS data. 
 
 For the best-fit model, with $\chi^2=\chi^2_{\mathrm{min}}$ , we  calculate the Probability-To-Exceed (PTE),  the probability of obtaining a higher $\chi^2$ value,
 \begin{equation}
\label{eqn:pte}
PTE=\int_{\chi^2_{\mathrm{min}}}^{\infty} \chi^2_m(x)dx,
\end{equation}
where $\chi^2_m$ is the $\chi^2$ distribution for $m$ degrees of freedom  \cite{Arrasmith:2017sek}. Unlikely events, or those in tension with theory given the experimental uncertainties, are signified by a low PTE. Consistently high PTEs might imply experimental uncertainties have been overestimated.

Finally, we  compute  the signal-to-noise ratio (SNR), inferred by assuming the signal is given by the best-fit theoretical model,
\begin{equation}
\label{eqn:snrth}
\mathrm{SNR}(\bar\tau)=\sqrt{\sum_{ij} \phat_{i,\mathrm{th}}(\bar\tau) \hat C_{ij}^{-1} \phat_{j,\mathrm{th}}(\bar\tau)}.
\end{equation}

\section{Analysis}
\label{sec:analysis}

\subsection{kSZ pairwise momentum results}
\label{sec:analysis:kSZ}

\begin{figure}[t!]
	\includegraphics[width=0.5\textwidth]{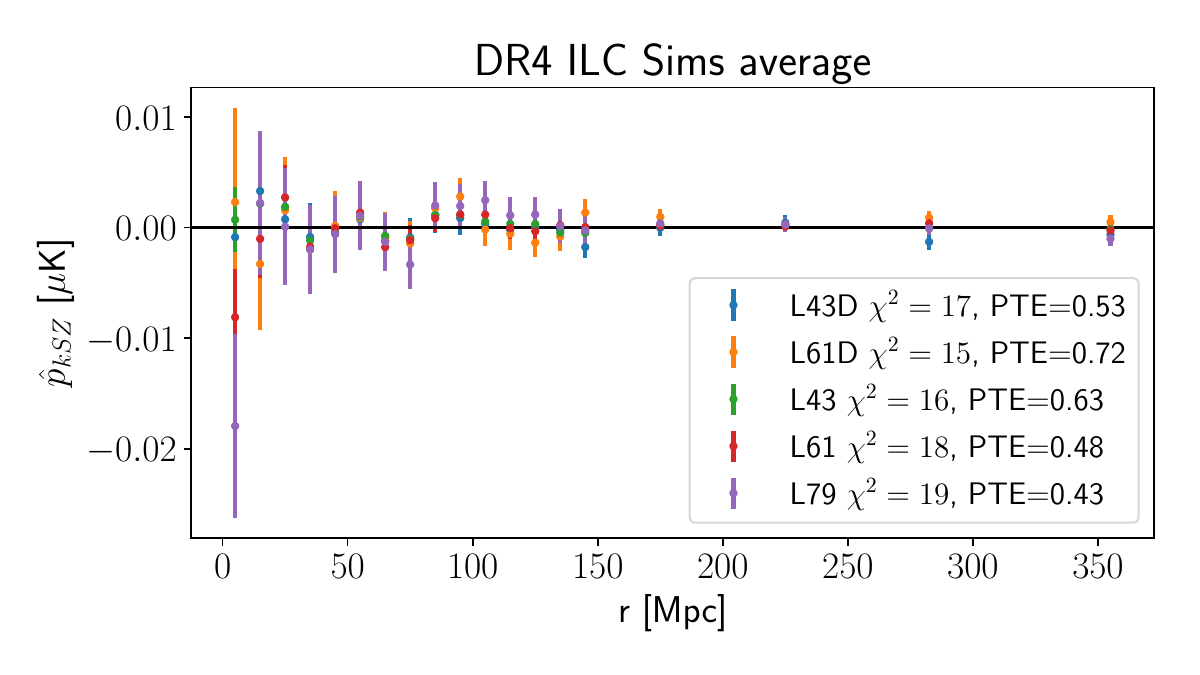}
	\includegraphics[width=0.5\textwidth]{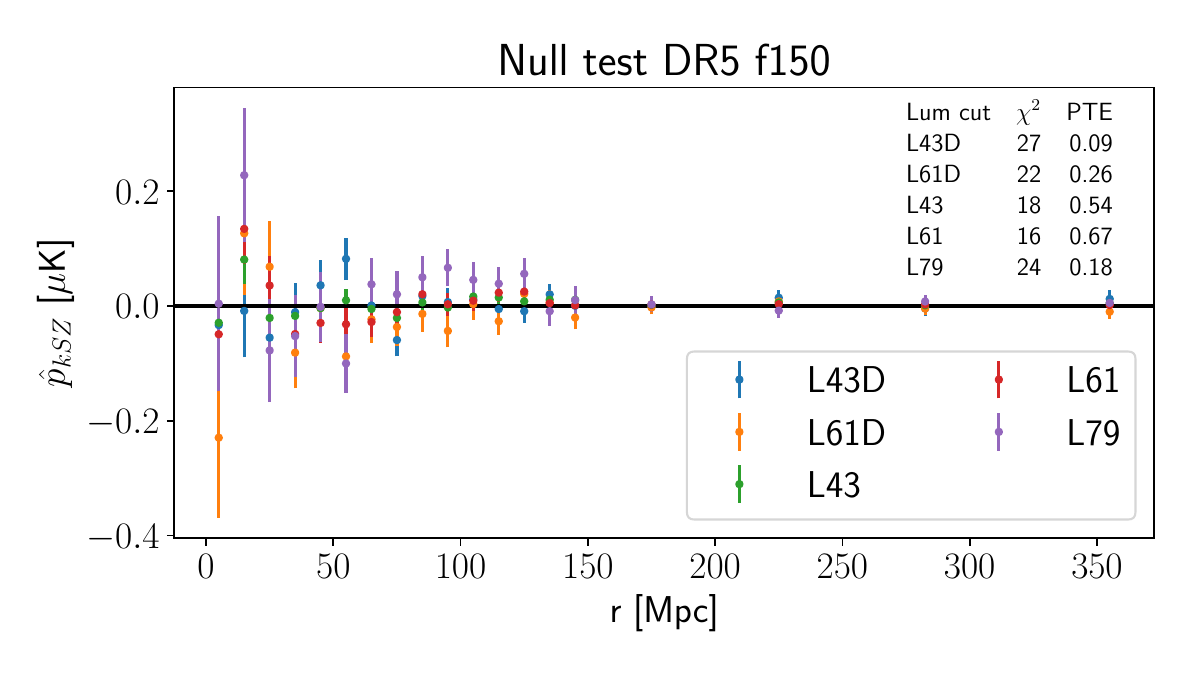}

	\caption{ Two null tests conducted on the pipeline for each galaxy tracer luminosity sample: [Upper] The mean and standard deviation of the pairwise estimator derived from the average sample covariance for the \cs 560 noise simulations. [Lower] A null test applied on the \caof map. Aperture photometries are taken from our science dataset for all luminosity bins, however the temperature decrement values are randomly shuffled which removes the pairwise signal. Error bars show one sigma uncertainty inferred with bootstrapping temperature decrements. The $\chi^2$ and the probabilities to exceed it are also given for each test. }
\label{fig:nulltests}
\end{figure}

The kSZ pipeline was tested using two methods respectively employing the \cs noise simulations and the \caof map. Firstly, we use the \cs noise simulations to apply the aperture photometry extraction, and compute the average pairwise kSZ signal and sample covariance over the 560 realizations. The second approach, testing the pipeline on the \caofbare, calculates the  aperture   temperature decrements for all galaxies in each luminosity-based tracer sample and then shuffles them while keeping the sky positions of the galaxies and redshifts fixed. The average and covariance of the resulting signals are calculated over 1,000 realizations.  Fig.~\ref{fig:nulltests}  shows the results of these two tests, which both show the expected effect, that the null tests remove the pairwise kSZ signal and leave a signal around zero with correlated uncertainties encapsulated in the covariance.

\begin{figure*}[!t]
	{
	\includegraphics[width=0.925\textwidth]{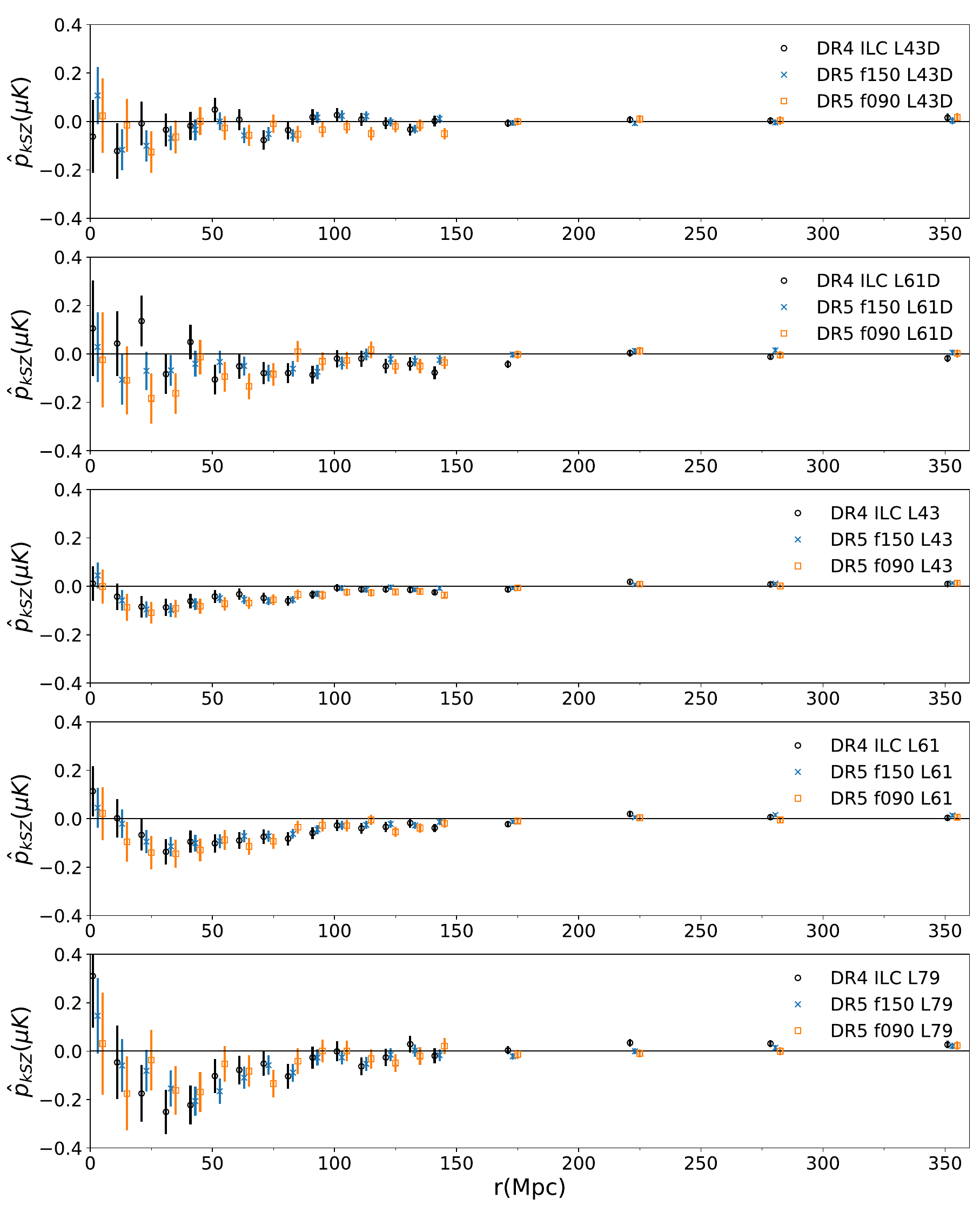}	
	}
\caption{Pairwise velocity correlations for the \cs [black circle] \caof [blue cross] and \can [orange square]   maps   for sources in the five luminosity-selected galaxy tracer samples:   [from top to bottom] $L43D$, $L61D$, $L43$, $L61$ and $L79$. Error bars show 1$\sigma$  bootstrap uncertainties. \vspace{0.125cm}
}
	\label{fig:mainres}
\end{figure*}

 In Fig.~\ref{fig:mainres} we present the pairwise kSZ measurements for   \csbare, \caof and \can   CMB maps with the SDSS DR15 luminosity-selected galaxy tracer samples along with the uncertainty estimates obtained from the bootstrap resampling of the same maps.
  
 The  $L43$, $L61$ and $L79$ samples  show a pairwise momentum profile with a negative signal amplitude reaching a maximum at separations around 25-50~Mpc. The negative amplitude is indicative of a  gravitational infall between the cluster pairs, with the mutual gravitational attraction falling off as one moves to cluster pairs separated by larger distances. The magnitude of the signal amplitude increases as the average luminosity of the sample increases, consistent with the observed clusters being more massive halos with larger optical depths, and with deeper gravitational potentials. At small scales, for $r<20$Mpc, the pairwise velocity correlation function has nonlinear contributions, and becomes positive, rather than negative as predicted by linear theory \cite{2012JCAP...11..009V, Okumura:2013zva,Sugiyama:2015dsa}. In the L43D and L61D disjoint bins, the kSZ signal is less discernible, consistent with an expectation that the signal in these groups/clusters  should be smaller since we expect their masses to be lower (coupled to the lower luminosities of the tracer galaxies), while the uncertainties, driven by their sample size, should be comparable to those in L79. We note that the three largest separation bins in the DR4 L79 data are positive, however these points are highly correlated (as shown in Fig.~\ref{fig:cov} in the Appendix) so that the deviation from null is not significantly anomalous.  
 
At each luminosity-selected galaxy tracer sample, the respective pairs of kSZ signals in Fig. \ref{fig:mainres} extracted from the three complementary maps show consistency within the 1$\sigma$ error bars. Given that each has a different approach to removing foreground emission, the consistency indicates that the results are robust against significant individual, distinct contamination from frequency dependent foregrounds. One concern that could arise is a potential impact of residual thermal SZ contamination in the signal especially from the most massive, luminous clusters. To address this we undertook two additional analyses. First, we compared the signal and covariance estimation for the \caof $L79$ sample, which has no upper luminosity limit, with that from a subsample with an upper luminosity threshold of $L<10^{11}L_{\odot}$ imposed. We find that there is no significant difference in the covariance estimates and no bias in the signal and the variation in the signal is at the level of a fraction of a standard deviation. Second, we analyzed a DR4 ILC map in which the tSZ signal has been deprojected \cite{2020PhRvD.102b3534M} and compare it to the \cs map analyzed in this paper. We find  no evidence of a bias in signal from the different tSZ treatments, and only that the noise is larger in the deprojected map.

\begin{table*}[!t]
  	\begin{tabular}{  | C{4.75em} |  C{5.5em} |C{3.0em} |C{3.0em} |C{3.0em} ||  C{5.5em} |C{3.0em} |C{3.0em} |C{3.0em} || C{5.5em} | C{3.0em} |C{3.0em} |C{3.0em} |}

\hline	
Tracer     &\multicolumn{4}{c||}{\cs }&\multicolumn{4}{c||}{\can } & \multicolumn{4}{c|}{\caof }
\\ \cline{2-13}   
sample&  $\bar\tau$  ($\times10^{-4}$)&$\chi^2_{\mathrm{min}}$ & PTE  & SNR	&  $\bar\tau$  ($\times10^{-4}$)&$\chi^2_{\mathrm{min}}$ & PTE  & SNR		& $\bar\tau$ ($\times10^{-4}$) &$\chi^2_{\mathrm{min}}$ & PTE  & SNR
  		\\ \hline
$L43D$	& 	0.18 $\pm$ 0.32	&	14	&	0.67	&	0.5		&	 0.83 $\pm$ 0.34  	& 12 		& 0.81	& 2.2   	& 	0.46 $\pm$ 0.24 	& 21	 & 0.24 	& 1.7	
		\\ 
$L61D$  	&  	0.69 $\pm$ 0.34	&	25	&	0.08	&	1.8		&	 1.07 $\pm$  0.35 	& 15 		&  0.59  	& 2.7 	&	 0.72 $\pm$ 0.26 	& 11	& 0.85	& 2.5	
		\\
$L43$ 	& 	0.47 $\pm$ 0.12	& 	22	& 	0.20	&	3.6	 	& 	0.65 $\pm$ 0.13   	& 13		&  0.71	 & 4.5  	& 	0.54 $\pm$ 0.09 	& 17 	& 0.42 	& 5.1
  		\\ 
$L61$ 	& 	0.74 $\pm$ 0.15	&	18	&	0.40	& 	4.4		& 	0.82 $\pm$  0.17	& 16		& 0.53	& 4.4		& 	0.69 $\pm$ 0.11 	& 10	& 0.92	& 5.4
  		\\  
 $L79$ 	& 	0.78 $\pm$ 0.23	&	21	&	0.21	&	3.0		& 	0.79 $\pm$ 0.27  	& 12		& 0.79  	& 2.6 	& 	0.88 $\pm$ 0.18 	&  13 & 0.76 	& 4.6	
 		\\ \hline
\multicolumn{13}{c}{}
		\end{tabular}  	
		\caption{The best-fit $\bar\tau$ estimates and 1$\sigma$ uncertainties  for the \cs [left], \can [center] and \caof [right] maps  for the five luminosity-selected galaxy tracer samples using the bootstrap uncertainty estimates. The corresponding $\chi^2$ (for 17 degrees of freedom),   SNR and PTE values are also given in each scenario.  }
  	\label{tab2}
  \end{table*} 

\begin{figure*}[!t]
	{
	\includegraphics[width=1.\textwidth]{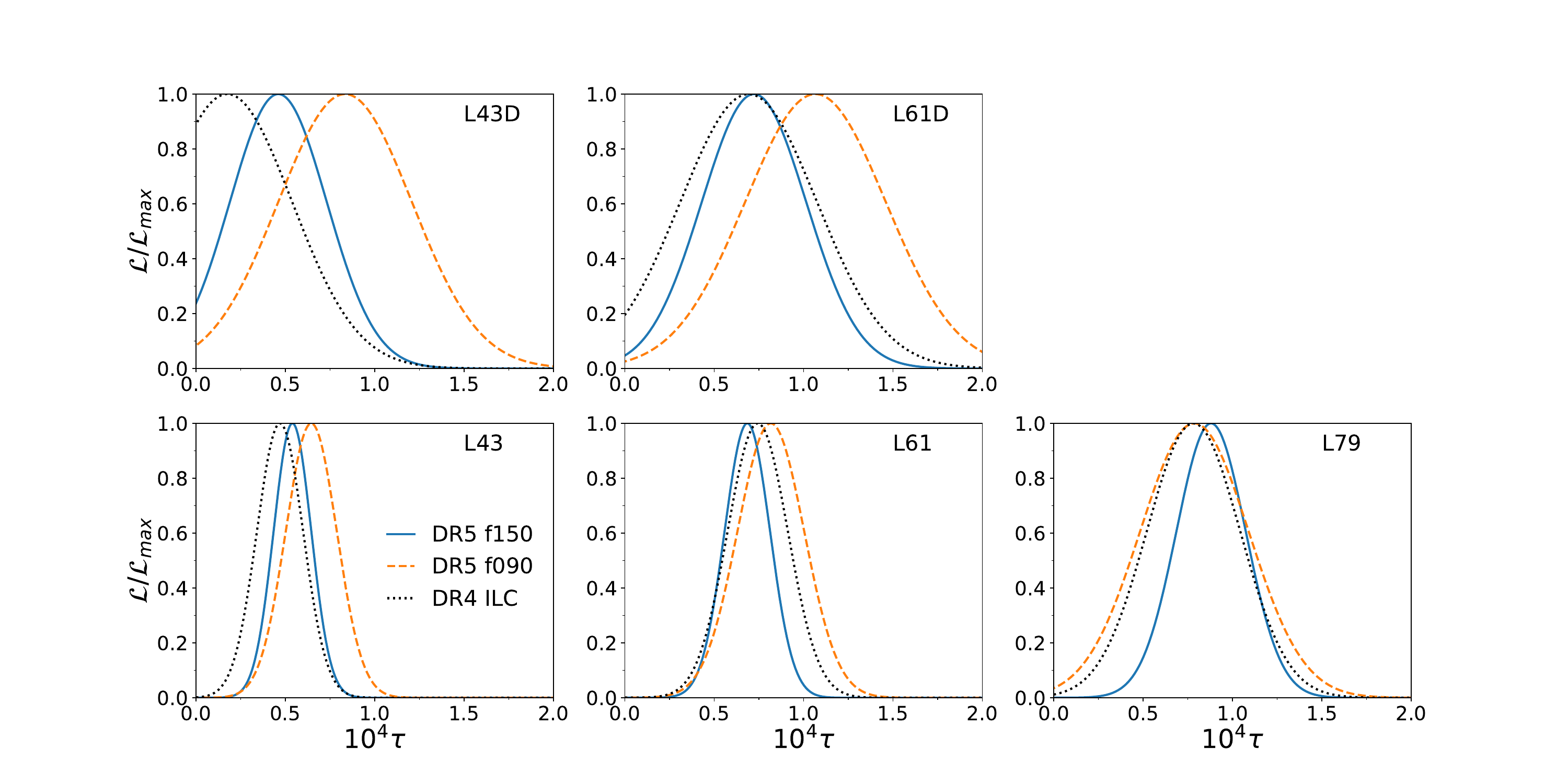}
	}
	\caption{The normalized likelihood for $\bar\tau$  estimates for the \caof   [blue, full],  \can [orange, dashed] and  \cs [black, dotted] maps  for each of the five luminosity-derived tracer samples  using the boostrap uncertainties.}
	\label{fig:Ltau}
\end{figure*}

  \subsection{Comparison with theoretical pairwise velocity predictions}
  \label{sec:tautheory}

We compare the observed pairwise correlations with theoretical linear pairwise velocity correlation predictions, given in (\ref{eq:vth}), for a {\it Planck} cosmology using the  code developed in  \cite{Mueller:2014nsa,Mueller:2014dba}. Using (\ref{eq:pksz}) we infer an effective measure of the cluster optical depth, $\bar\tau$, for the observed samples using 17 spatial separation bins spanning 20~Mpc$<r<$395~Mpc. We exclude $r< 20$~Mpc as at these scales nonlinear velocity effects become significant that are not incorporated in the linear theoretical fit.

In Table~\ref{tab2} we compare the optical depth constraints for the \csbare, \caof and \can maps, using  bootstrap estimates  as directly obtained from the maps. The probability that a $\chi^2$ would exceed the minimum, best-fit $\chi^2_{\mathrm{min}}$ (PTE)  and the signal-to-noise ratio (SNR) for the best-fit scenario are also given.

For each map, the average optical depth obtained in each sample increases with increasing mean host halo mass in each bin (and the luminosity of the  LRG used as the group/cluster center tracer). The  uncertainties in the optical depth measurements increase in tandem with the 
signal uncertainties, principally driven by the number of galaxies in the luminosity bin.  In Appendix~\ref{sec:app:covest}, we summarize how the impact on $\bar\tau$ fits of using the covariances derived from the JK and bootstrap methods, and from the dispersion in signals obtained from the 560 noise sims.

Our best measured detections of $\bar\tau$ are in the \caof map with the L61 luminosity cut, for which we obtain SNR of 5.4 with the derived best-fit mass-averaged optical depth of ${\bar\tau}=(0.69\pm 0.11)\times 10^{-4}$ with a $\chi^2$ of 10 for 17 degrees of freedom.

We assess the sensitivity of these results to uncertainties in the assumption about the minimum and maximum halo mass, $M_{\mathrm{min}}$ and $M_{\mathrm{max}}$ in (\ref{eq:bq}) using the \caof map. The bin with the highest mass tracer sample, L79,  is the sample for which the maximum mass sensitivity would be most pronounced. For the default maximum mass of 10$^{16}M_{\odot}$ we obtain   $\bar\tau$ = (0.88 $\pm$ 0.18)$\times 10^{-4}$. If we reduce the maximum mass by an order of magnitude, to $10^{15}M_{\odot}$, we find that the change in the $\bar\tau$ constraints is small, with no change to the best-fit value, a small reduction to the uncertainties $\bar\tau$ = (0.88 $\pm$ 0.17)$\times 10^{-4}$ and a change of 0.1 in the $\chi^2$. The small impact sensitivity to $M_{\mathrm{max}}$ is expected, as this  truncates the high mass tail of the halo mass function which only contributes a small fraction of the halos over which the mass averaged optical depth is calculated. By corollary, we anticipate that the minimum mass will have a larger impact since it impacts more of the halos. Halving the minimum mass lowers the halo bias parameters in the mass averaged correlation function, and reduces the amplitude of predicted pairwise velocity signal. This, in turn, requires  a larger $\bar\tau$ estimate to fit the theoretical velocity prediction to the pairwise momentum data.  The variation  introduced by a factor of two theoretical uncertainty  in the minimum halo mass is found to be subdominant to the experimental uncertainties: for the \caof $L43$ sample, using the  default minimum mass of $M=0.52 \times 10^{13}M_{\odot}$ we obtain $\bar\tau=(0.54\pm 0.09)\times 10^{-4}$. For a minimum mass of half that size we find $\bar\tau = (0.57\pm 0.10)\times10^{-4}$ and when doubled, $\bar\tau = (0.50\pm0.09)\times10^{-4}$.  The $\chi^2$ fit changes by $< 0.1$, varying between 17.4 and 17.6, and the SNRs remain unchanged.   

 In Fig.~\ref{fig:Ltau} we present the likelihoods for the effective $\bar\tau$ value for these three maps. The figure shows how the best-fit values of $\bar\tau$ are consistent across the three maps, and show an increase as the minimum luminosity  threshold  for the galaxy tracer sample increases, congruent with an increase in the integrated line of sight number density of electrons  with halo mass. Again consistent with the decrease in the sample size, the uncertainty in the $\bar\tau$ measurement  increases when one considers sequentially higher luminosity thresholds. 
 
 The consistency in the signals between the 90GHz, 150GHz and those from the component separated map (across which the tSZ contributions vary)  indicates that residual tSZ contamination is not significant. Fig~3. shows that the $\tau$ estimates obtained from each of the 3 maps for the L79 sample, focused on the most massive clusters with the largest potential residual foregrounds, are very consistent. Similarly, the L61 cumulative bin, which doesn't have an upper luminosity bound, and the L61D bin, which does, show consistent $\tau$ estimates, though the L61D results have a greater variance due to the smaller sample size, showing that uncertainties in the high luminosity clusters are not biasing the results. 

\begin{figure*}[!t]
	{
	\includegraphics[width=\textwidth]{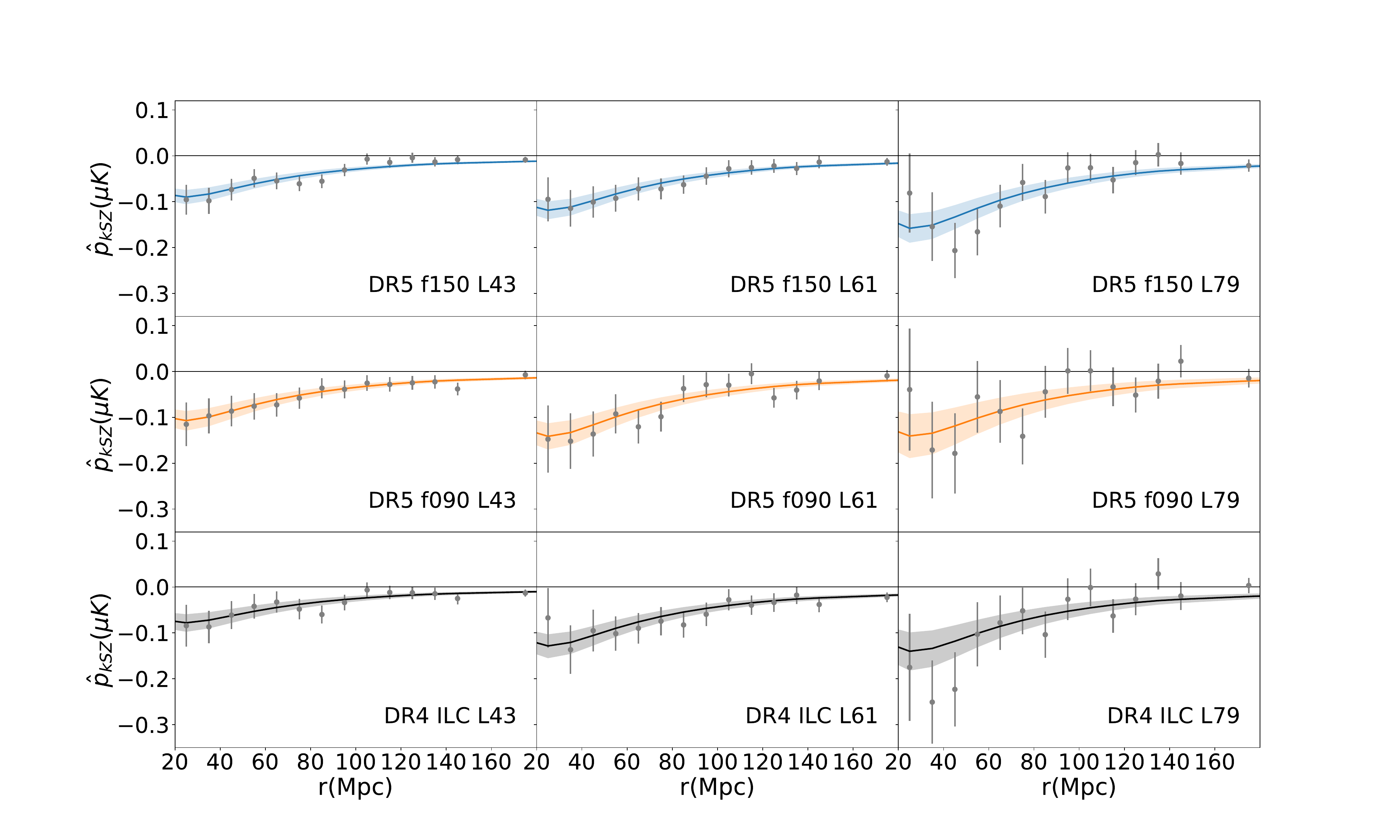}	
	}

	\caption{The extracted pairwise signal for the \cs [black, lower] and \can [orange, middle] and \caof [blue, upper] maps for the three cumulative luminosity-selected galaxy tracer samples, $L43$ [left], $L61$ [center], $L79$ [right], overlaid with the  theoretical pairwise velocity model using the {\it Planck} best-fit cosmology corresponding to the best-fit $\bar\tau$ value and 1$\sigma$ boostrap-derived uncertainties.}	
	\label{fig:phattau}
\end{figure*}

 In Fig.~\ref{fig:phattau}, we overlay the observed pairwise correlations for the \caofbare, \canbare, and \cs  maps,  with the theoretical models using the best-fit and $1\sigma$ constraints on $\bar\tau$ and the {\it Planck} cosmology pairwise velocity predictions.

\section{Conclusions}
\label{sec:conclusions}

In this work, we present measurements of the optical depth of clusters derived from cross-correlations of pairwise kSZ effect for  three co-added maps, \caof and \canbare, utilizing the most recent ACT DR5 data combined with {\it Planck} PR2 and PR3 data, and a component separated map, \csbare, using ACT DR4 and {\it Planck} PR2 data. The kSZ signal is obtained by correlating the maps with the SDSS DR15 galaxy catalog using  luminous red galaxies as tracers of the group/cluster center  in five luminosity cuts with a minimum luminosity threshold of $4.3 \times 10^{10}L_{\odot}$.

We use bootstrap-derived estimates, derived from each map, for the covariance used to derive kSZ estimates of the mass averaged optical depth for the samples.
As detailed in Appendix~\ref{sec:app:covest}, we use the bootstrap method after comparing covariance estimates from bootstrap and jackknife methods, and those from variances across noise sims. We find that JK uncertainties systematically overestimate  the uncertainties while the bootstrap method more closely aligns with those from the averaged noise sims.

Using the bootstrap uncertainties,  the highest SNR is obtained for the L61 tracer sample and \caof map, with a 5.4$\sigma$ measurement for the best-fit theoretical model relative to the null signal, with a mass-averaged optical depth of ${\tau}=(0.69\pm 0.11)\times 10^{-4}$.

In this multi-frequency analysis, we find consistent results for the component separated map, \csbare, and the co-added, \can and \caofbare, maps implying a robustness of the signals extracted, and $\bar\tau$ fit estimates, to potential frequency dependent contamination.  

A number of refinements to the aperture photometry method have been implemented to improve the precision of signal estimation relative to previous work. This includes considering assumptions on pixel size, pixel reprojection, galaxy versus pixel centering of CMB submaps, fractional pixel weighing, and noise-weighting, as outlined in Appendix \ref{sec:comparison}. We leave to future work the potential impact of the brightest galaxy being displaced from the cluster center.

In a companion paper, \evepaper, the kSZ $\bar\tau$ results obtained here are compared with those derived from thermal SZ measurements from the same data and  theoretical predictions based on the cluster baryon content using a Navarro-Frenk-White  profile \citep{Navarro:1995iw}.

 This analysis paves the way for pairwise kSZ work with this pipeline on upcoming and future data, including upcoming CMB instruments, for example, Simons Observatory \cite{Ade:2018sbj}, SPT-3G \cite{Benson:2014qhw}, CMB-S4 \cite{Abazajian:2016yjj} and the FYST telescope \cite{Aravena:2019tye}, in tandem with upcoming spectroscopic and photometric large scale structure surveys including the Dark Energy Spectroscopic Instrument (DESI) \citep{Levi:2013gra}, the ESA-NASA Euclid Telescope \cite{Amendola:2016saw}, the Vera Rubin Observatory  \citep{Abell:2009aa,Abate:2012za} and the Nancy Roman Space Telescope \citep{2019arXiv190205569A,Eifler:2020vvg}.

\section*{Acknowledgments}
VC and RB acknowledge support from DoE grant DE-SC0011838, NASA ATP grant 80NSSC18K0695, NASA ROSES grant 12-EUCLID12-0004 and funding related to the Roman High Latitude Survey Science Investigation Team.  EMV acknowledges support from the NSF Graduate Research Fellowship Program under Grant No. DGE-1650441. NB acknowledges support from NSF grant AST-1910021, NASA ATP grant 80NSSC18K0695 and from the Research and Technology Development fund at the Jet Propulsion Laboratory through the project entitled ``Mapping the Baryonic Majority''. EC acknowledges support from the STFC Ernest Rutherford Fellowship ST/M004856/2 and STFC Consolidated Grant ST/S00033X/1, and from the European Research Council (ERC) under the European Union’s Horizon 2020 research and innovation programme (Grant agreement No. 849169). SKC acknowledges support from NSF award AST-2001866. JD is supported through NSF grant AST-1814971.  RD thanks CONICYT for grant BASAL CATA AFB-170002.  RH acknowledges funding from the CIFAR Azrieli Global Scholars program and the Alfred P. Sloan Foundation.  JPH acknowledges funding for SZ cluster studies from NSF grant number
AST-1615657. KM acknowledges support from the National Research Foundation of South Africa.  MDN acknowledges support from NSF CAREER award 1454881. CS acknowledges support from the Agencia Nacional de Investigaci\'on y Desarrollo (ANID) through FONDECYT Iniciaci\'on grant no. 11191125. ZX is supported by the Gordon and Betty Moore Foundation. 
 
This work was supported by the U.S. National Science Foundation through awards AST-0408698, AST-0965625, and AST-1440226 for the ACT project, as well as awards PHY-0355328, PHY-0855887 and PHY-1214379. Funding was also provided by Princeton University, the University of Pennsylvania, and a Canada Foundation for Innovation (CFI) award to UBC. ACT operates in the Parque Astron\'omico Atacama in northern Chile under the auspices of the Comisi\'on Nacional de Investigaci\'on Cient\'ifica y Tecnol\'ogica de Chile (CONICYT). 

  Canadian co-authors acknowledge support from the Natural Sciences and Engineering Research Council of Canada. The Dunlap Institute is funded through an endowment established by the David Dunlap family and the University of Toronto.
  
Funding for the Sloan Digital Sky 
Survey IV has been provided by the 
Alfred P. Sloan Foundation, the U.S. 
Department of Energy Office of 
Science, and the Participating 
Institutions. 
SDSS-IV acknowledges support and 
resources from the Center for High 
Performance Computing  at the 
University of Utah. The SDSS 
website is www.sdss.org.
SDSS-IV is managed by the 
Astrophysical Research Consortium 
for the Participating Institutions 
of the SDSS Collaboration including 
the Brazilian Participation Group, 
the Carnegie Institution for Science, 
Carnegie Mellon University, Center for 
Astrophysics | Harvard \& 
Smithsonian, the Chilean Participation 
Group, the French Participation Group, 
Instituto de Astrof\'isica de 
Canarias, The Johns Hopkins 
University, Kavli Institute for the 
Physics and Mathematics of the 
Universe (IPMU) / University of 
Tokyo, the Korean Participation Group, 
Lawrence Berkeley National Laboratory, 
Leibniz Institut f\"ur Astrophysik 
Potsdam (AIP),  Max-Planck-Institut 
f\"ur Astronomie (MPIA Heidelberg), 
Max-Planck-Institut f\"ur 
Astrophysik (MPA Garching), 
Max-Planck-Institut f\"ur 
Extraterrestrische Physik (MPE), 
National Astronomical Observatories of 
China, New Mexico State University, 
New York University, University of 
Notre Dame, Observat\'ario 
Nacional / MCTI, The Ohio State 
University, Pennsylvania State 
University, Shanghai 
Astronomical Observatory, United 
Kingdom Participation Group, 
Universidad Nacional Aut\'onoma 
de M\'exico, University of Arizona, 
University of Colorado Boulder, 
University of Oxford, University of 
Portsmouth, University of Utah, 
University of Virginia, University 
of Washington, University of 
Wisconsin, Vanderbilt University, 
and Yale University.

\appendix

\section{Impact of analysis assumptions on kSZ signal extraction}
\label{sec:comparison}

In this section we present a study of the impact of  various  analysis assumptions made in this paper, as presented in section \ref{sec:analysis}.  As part of this we  assess the impact of differences in the assumptions in this paper and in the previous ACT kSZ analysis,  DB17, in which  the first three seasons of ACT data \cite{Das:2013zf} and the first season of ACTpol data \cite{Naess:2014wtr} were cross-correlated with  clusters identified through a color-luminosity selected SDSS DR11 galaxy sample.

\begin{table*}[!t]
\begin{tabular}{|c|c|c|l|l|l|l|l|}
\hline
{\bf Scenario} 	& ACT   	& SDSS   	& Galaxy query 	& CDELT 	& Submap    & Aperture  & Submap    \\ 
 			& map 	& sample    				&  \& K-correction 	& &rounding &  averaging   		& centering \\ \hline
{\bf 1}       &  DB17& DR11  & As in DB17 & Approx.     & Approx.                 & Full        & Pixel                                     \\ \hline
{\bf 2}       &  DB17& DR11  & As in DB17&    \textcolor{blue}{{\bf Precise}}         & Approx.                    & Full         & Pixel          \\ \hline
{\bf 3}        & DB17 & DR11 & As in DB17 & Precise         &  \textcolor{blue}{{\bf Precise}}                       & Full            & Pixel          \\ \hline
{\bf 4}        &  \textcolor{blue}{{\bf  \cs}} & DR11& As in DB17 & Precise         & -- &  \textcolor{blue}{\bf Fractional }&  \textcolor{blue}{\bf Galaxy }          \\ \hline
{\bf 5}        &\cs  &   \textcolor{blue}{{\bf DR15}} & \textcolor{blue}{\bf \evepaper }  & Precise         & --                        & Fractional   & Galaxy        \\ \hline
{\bf 6}        &  \textcolor{blue}{\bf\caof}  &  DR15 &  \evepaper   & Precise         & --                        & Fractional        & Galaxy        \\ \hline
\end{tabular}
  	\caption{Summary of scenarios utilized in Figure \ref{fig:stepbystep} which demonstrate in a step-by-step fashion, with step changes highlighted in blue, the impact of various assumptions used in this analysis and the earlier analysis in DB17. }
  	\label{tab4}
\end{table*}
 
\begin{figure}[t!]
	{
	\includegraphics[width=0.50\textwidth]{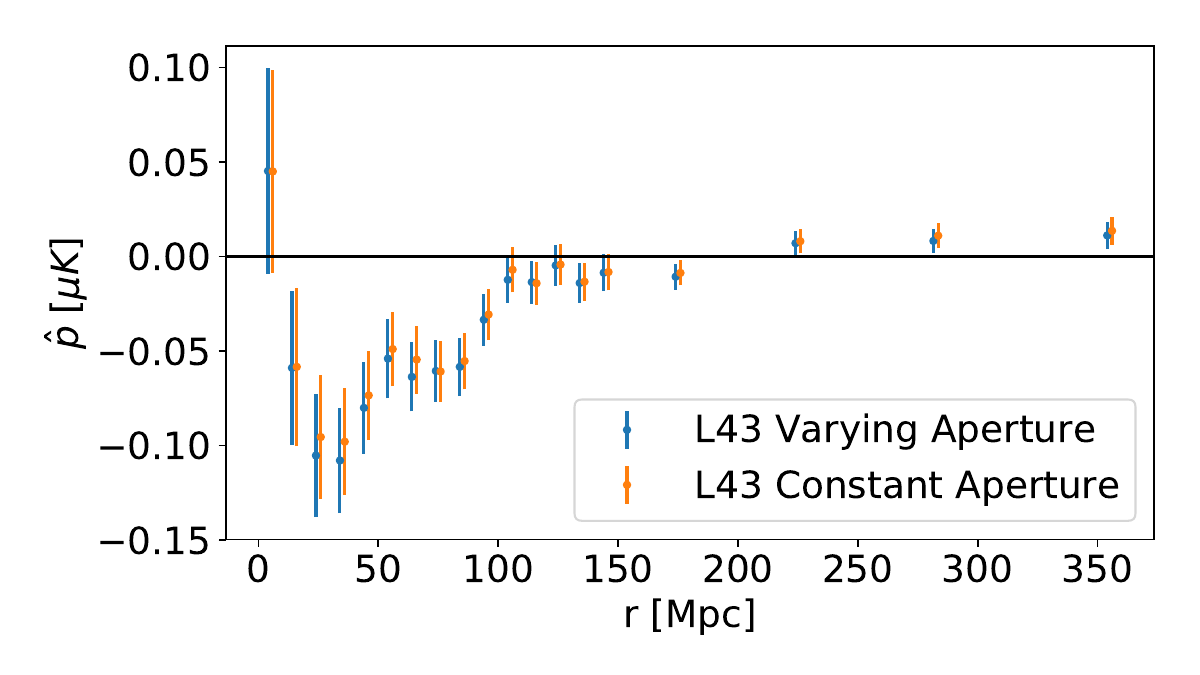}
	}
	\caption{Comparison of the impact of a constant [red] versus redshift-varying [blue] aperture size  in the pairwise momentum estimation, $\phat$, for the \cs map and the DR15 $L43$  sample.  }
	\label{fig:aperture}
\end{figure}

We consider the impact of six specific  analysis assumptions.
  \begin{itemize}
\item{\it Galaxy sample selection:} As outlined in \ref{sec:data:lss}, we obtain the galaxy sample using the SDSS SQL queries and a K-correct code,  described  in detail in \evepaper. 
   
    \item {\it Aperture photometry -- Pixel size and submap precision:} We found that two pixel parameters used in the DB17 analysis to determine which pixels are included in the aperture photometry for a given cluster  were approximated/rounded: the ``CDELT" parameter determining the pixel size and a second parameter determining which pixel serves as the central pixel for the submap. We found that using the full rather than approximated values led to differences in which pixels are identified in the disc/annulus and a consequent change in the predicted signal.
    \item {\it Aperture photometry -- Reprojection:} We account for the geometrical projection effects that modify the equal area treatment in aperture photometry when the cluster location is near the poles rather than the equator  by using the \texttt{Pixell} reprojection subroutine.   
\item{\it Aperture photometry -- Pixel vs. tracer galaxy-centering:} In DB17, the aperture photomety was centered around  the center of the pixel in which the tracer galaxy is located. In this analysis, we implement a ``galaxy-centered" approach rather than a ``pixel-centered" approach in translating the coordinates. This means we determine which temperature values are within the disc and annulus centered on the coordinates of the tracer galaxy itself, as opposed to translating the tracer galaxy coordinates to the reference pixel  and populating the disc based on the center of the pixel.  As in DB17, we use postage stamps rather than full map to speed up the code. \texttt{Pixell} is used to create reprojected postage stamps recentered on the galaxy location.

  %
  \begin{figure}[!t]
      \centering
      {\includegraphics[width=0.475\textwidth]{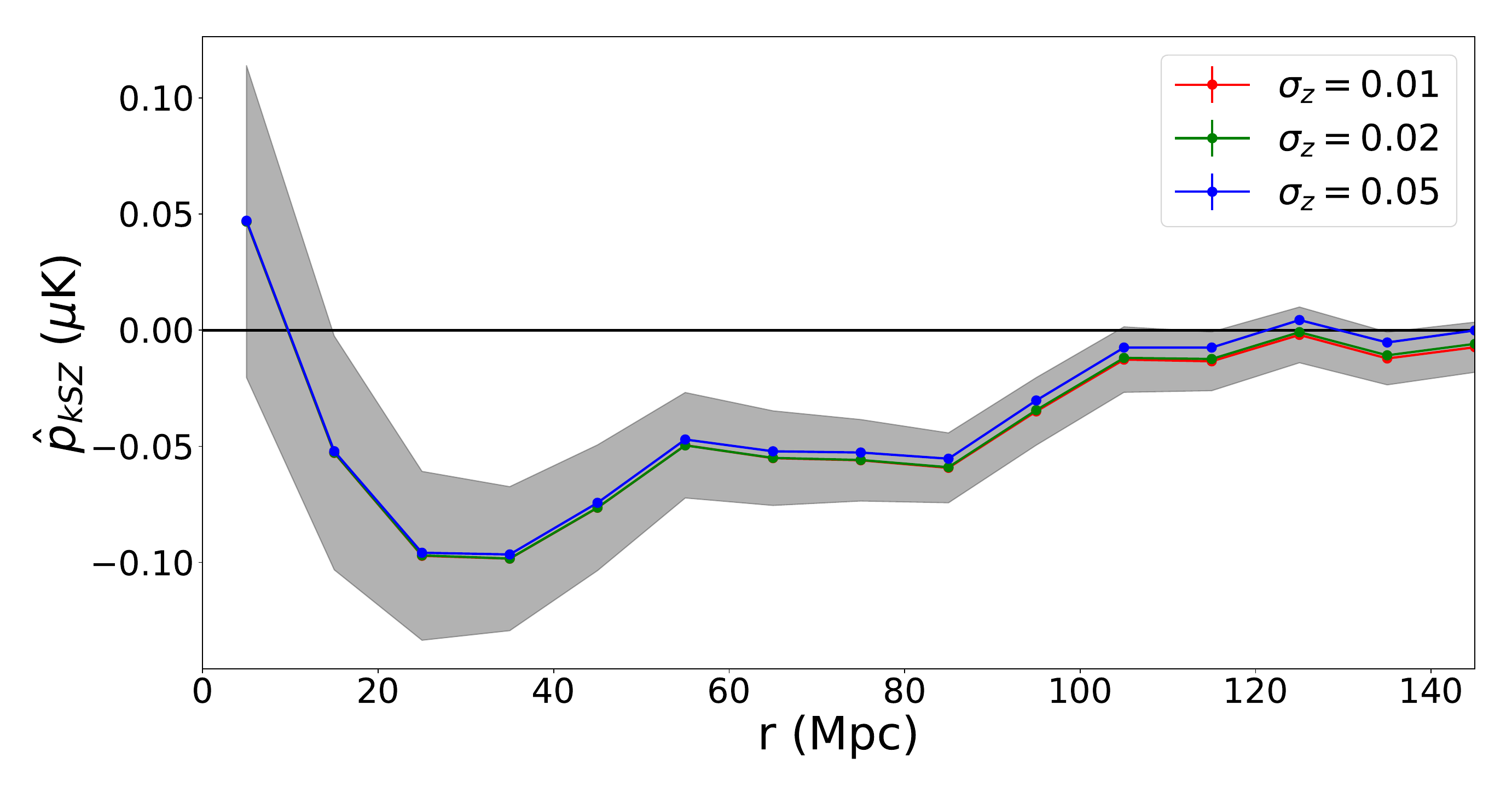}}
      {      \includegraphics[width=0.45\textwidth]{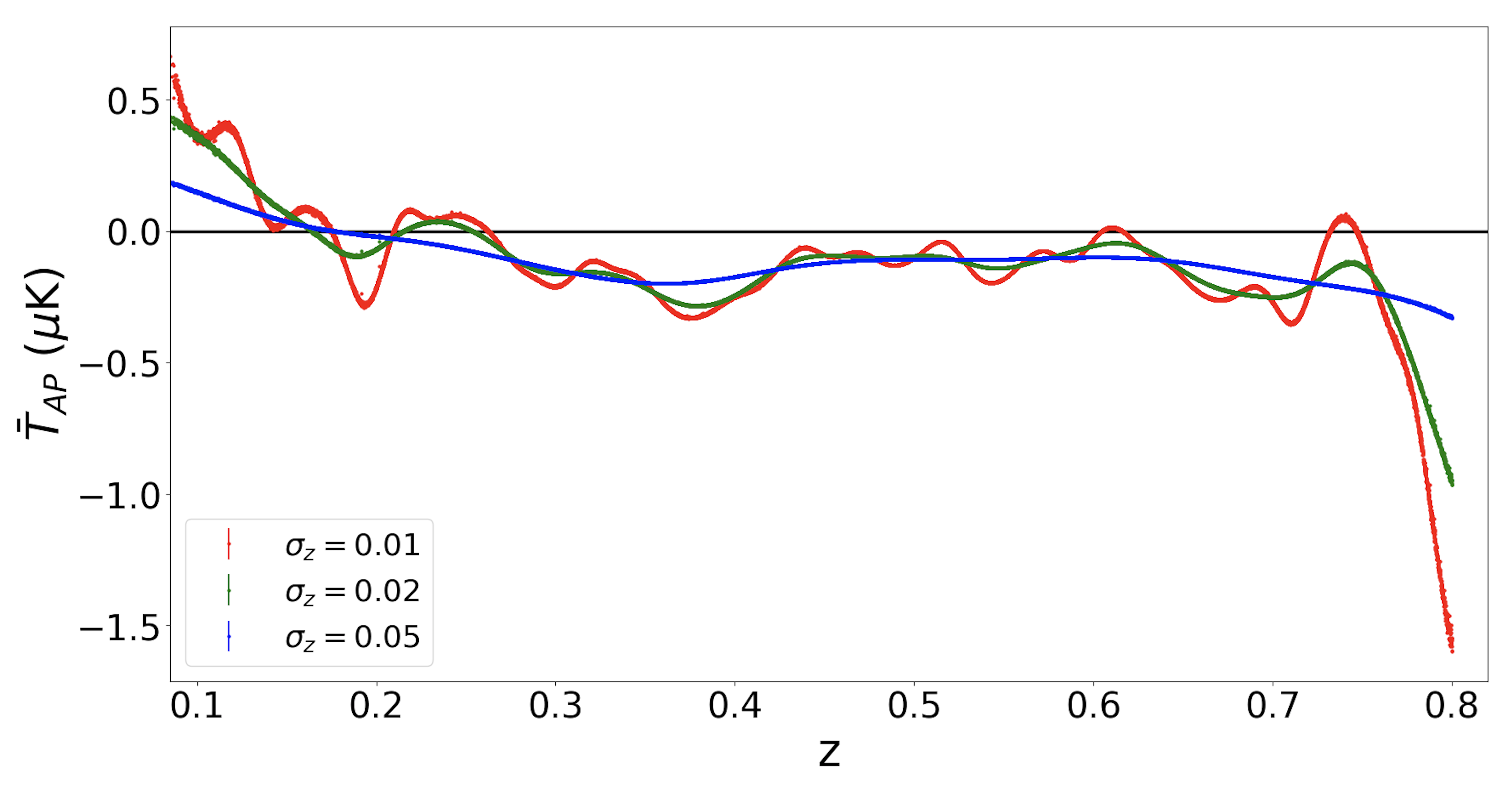}}
      \caption{Comparison of the impact of the redshift smoothing factor in the aperture photometry estimation for the \cs map and the DR15 $L43$  sample. [Upper] The uncertainties introduced in $\hat p_{\mathrm{kSZ}}$ when varying $\sigma_z$=0.01, 0.02, 0.05 are compared with 1$\sigma$ uncertainties from noise simulations (assuming $\sigma_z=$ 0.01). [Lower] The individual $\bar{T}_{AP}$ realizations for each case.}
      \label{fig:wiggleT}
  \end{figure}

\item{\it Aperture photometry -- Fractional pixel weighting:} In implementing the aperture photometry, we account for cases in which pixels are only partially included in the annulus or disc. This includes both pixels centered outside of the disc or ring, and is especially important for pixels that span between the disc and annulus. This is done by creating a finer resolution pixel grid as described in section~\ref{sec:form:pw}.
\item{\it Aperture photometry -- Noise weighting:} We compared  flat and noise-weighting schemes for differencing the kSZ temperature decrements in the pairwise momentum estimator in (\ref{eq:phat}) and found little difference in the resulting signal. 
\item{\it Aperture photometry -- Aperture size:} We compare the signal and covariance derived from aperture photometry with a fixed 2.1' aperture at all redshifts, and an aperture that varies with redshift in proportion to the angular diameter distance, $D_A$, calculated assuming the best fit Planck cosmology, $\Theta(z) = 2.1' D_A(z)/D_A(z=0.5)$, scaled to be 2.1' at the sample's mean  redshift. In Fig.~\ref{fig:aperture} we show that the signal and covariance, for the tracer galaxy sample used in this analysis, are minimally affected by the choice of fixed or redshift-varying aperture size.

  \item {\it Pairwise estimation -- Mitigating redshift evolution: } In both this paper and DB17, the effects of redshift evolution in the kSZ signature within a comoving separation bin are accounted for by subtracting a redshift averaged temperature estimate, $\bar T_{AP}$,  in (\ref{eq:TAPbar}). In Fig.~\ref{fig:wiggleT} we show that the assumptions about the redshift smoothing factor $\sigma_z$ do not significantly impact the signal extraction, with the differences in signal being far smaller than the statistical uncertainties in the covariance estimation process.
\end{itemize}

\begin{figure}[t!]
	{
	\includegraphics[width=0.50\textwidth]{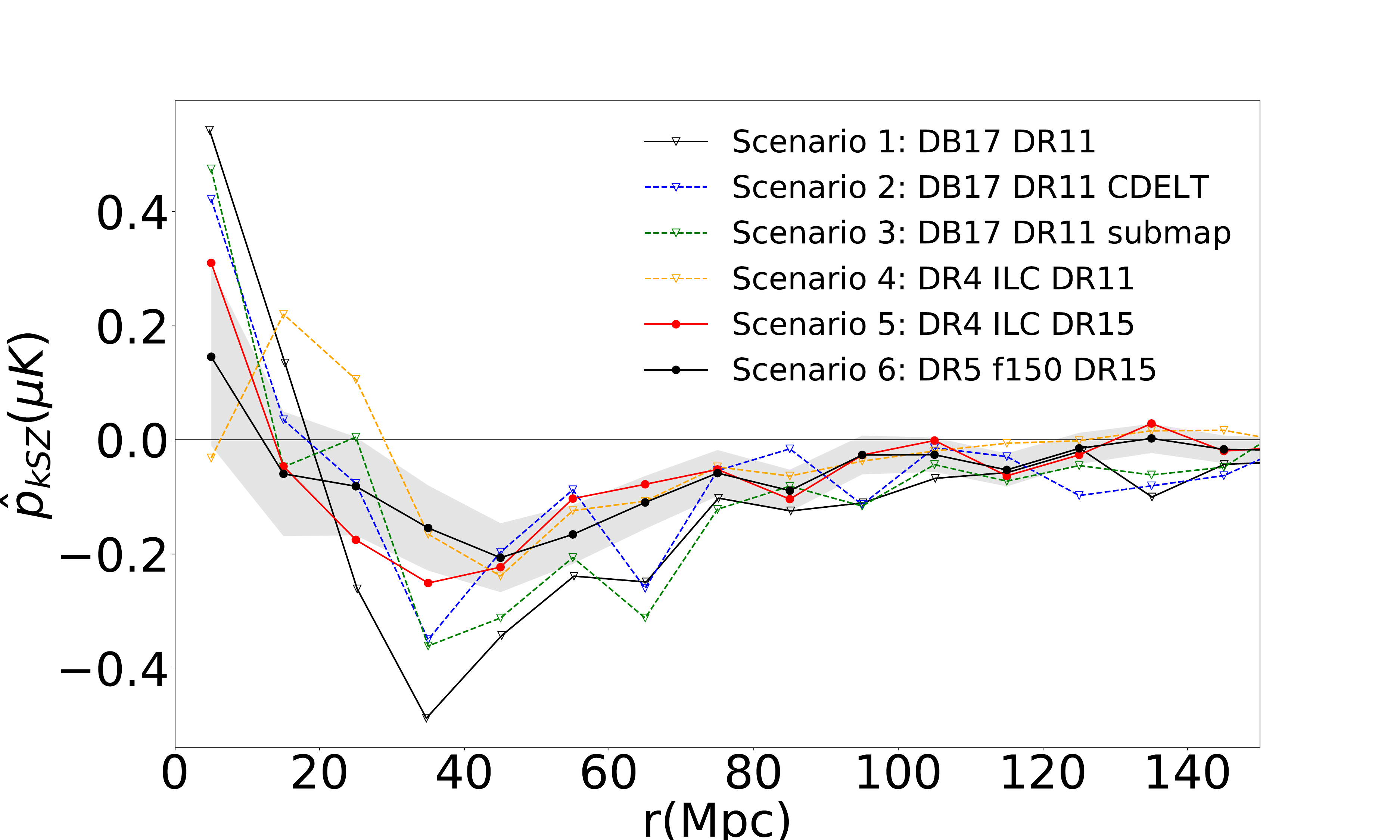}
	}
	\caption{Comparison of the impact of step-by-step changes in the aperture photometry assumptions in pairwise momentum estimation, $\phat$ obtained from those used in  DB17 to this work.  The stepwise changes in analysis and data are given in Table ~\ref{tab4} for samples with a $L>7.9 \times 10^{11}L_{\odot}$ luminosity cut. The gray shaded region shows the 1$\sigma$ boostrap-derived uncertainties for the \caof analysis.}
	\label{fig:stepbystep}
\end{figure}

In Table \ref{tab4} and Fig.~\ref{fig:stepbystep} we present six scenarios that allow stepwise comparison of the kSZ pairwise signal obtained in DB17 to that obtained in our main results when one factors in various updates to the analysis approach. The starting point, scenario 1, utilizes the DB17 CMB map and DR11 sample of 20,000 galaxies and a luminosity cut of $L>7.9\times10^{10}L_\odot$.  The end point, scenario 6, uses the \caof map and L79 DR15 sample of 103,159 galaxies used in the analysis in this paper.  %
The transition from scenario 1 to scenario 2 shows that using the precise value of the CDELT parameter in the aperture photometry temperature determination reduces the peak signal by $\sim40\%$ at $r=35$~Mpc/h.
Scenarios 2 and 3 show the effect  of rounding the pixel size and submap centering parameters is less pronounced but does still create a variation in the recovered pairwise signal. 
Scenarios 3 and 4 show the impact of changing from the DB17 map to the component separated map, \csbare, and introducing a scheme in which the aperture photometry is centered on the galaxies themselves, not the pixel center in which the galaxy resides and in which the temperatures include fractional weighting of pixels that overlap the edges of the apertures.
The comparison of scenarios 4 and 5 shows the impact of the transition from the 20,000 DR11 galaxy sample to the  57,828 galaxies in the DR15 sample that overlap with the \cs map.  The parallelized \texttt{Pixell} Python code used in this paper is also employed. The peak amplitude shifts slightly to fall between 35 to 45~Mpc, slightly larger cluster separation than in the  DB17 signal.
The final transition, from scenario 5 to scenario 6, shows the difference in signal extraction between \cs and \caof maps for the L79 DR15 sample.  The peak signal shifts towards a separation of 45~Mpc while the amplitude of the peak signal changes only slightly relative to the previous changes. In comparing scenarios 1 and 6, the changes in combination lead to a reduction in the  kSZ signal measured in our analysis relative to that in DB17.

\section{kSZ pairwise momentum covariance estimation}
\label{sec:app:covest}

In this section we compare the covariances of the kSZ pairwise momentum estimates 
using
 jackknife (JK) and bootstrap techniques
 for the three maps (\cs, \caof and \can) and, for the \cs map, also compare them to the estimate obtained by averaging over many simulated maps.

For the JK estimation,  the clusters in the luminosity bin being considered are binned into $N$ subsamples, removing each subsample exactly once, and each time computing the pairwise estimator according to the remaining ($N-1$) subsamples. 
The covariance matrix is then given by
\begin{equation}
\label{eq:JK}
\hat C_{ij,JK}= \frac{N-1}{N}\sum_{\alpha=1}^{N} (\hat p_i^\alpha - \bar p_i) (\hat p_j^\alpha - \bar p_j),
\end{equation} 
where $\hat p_i^\alpha$ is the signal extracted from the $\alpha^{\mathrm{th}}$ JK sample  for the $i^{\mathrm{th}}$ separation bin, and $\bar p_i$ is the mean of the $N$ JK samples
 \cite{2016MNRAS.461.3172S}.   The  inverse  of $\hat C_{ij,JK}$ is a biased estimator of the true inverse covariance, and to address this one uses an additional correction factor \cite{Hartlap:2006kj},
 \begin{equation}
\label{eq:JKinv}
 \hat{C}_{ij}^{-1}=  \frac{(N-K-2)}{(N-1)}\hat{C}^{-1}_{JK,ij} 
\end{equation} 
where $K$ is the number of comoving separation bins used in the analysis. For our analysis $N=1,000$ and $K=19$ separation bins in total, although we use $K=17$ in the performing the optical $\bar\tau$ fits (excluding the two smallest separation bins), as described in section~\ref{sec:tautheory}.

For the bootstrap estimation, we randomly reassign the temperature decrements of galaxy positions allowing for repeated values (sampling with replacement). We repeat this process 1,000 times, computing the pairwise kSZ estimator for each replicant sample. We compute the covariance matrix as the sample covariance of the list of pairwise kSZ curves obtained with this process. We note that while the effects of filter overlapping have been shown to cause bootstrap errors to underestimate the covariance for large apertures, the effect has been shown to be negligible for the smaller aperture size used in this analysis (\cite{Schaan:2020qhk}, Appendix D and Fig.~27).

For the \cs we  have access to 560  simulated maps  produced in \cite{Darwish:2020fwf}, which include primary CMB, lensing, and Gaussian but spatially inhomogeneous extragalactic foregrounds and noise  due to detector correlations and scan strategy, as generated using the pipeline described in \cite{Choi:2020ccd} (specifically, we use simulation version $v1.2.0$). 
Equivalent simulated
 realizations are not, however, available for the \caof and \can maps.   We  use one 
 simulated map (to keep computation times bounded), to compute JK or bootstrap uncertainties using 1,000 random catalog resamplings and compare these estimates to the average sample covariance obtained from the 560 simulations.


\begin{figure}[t!]
    \centering
    	\includegraphics[width=\columnwidth]{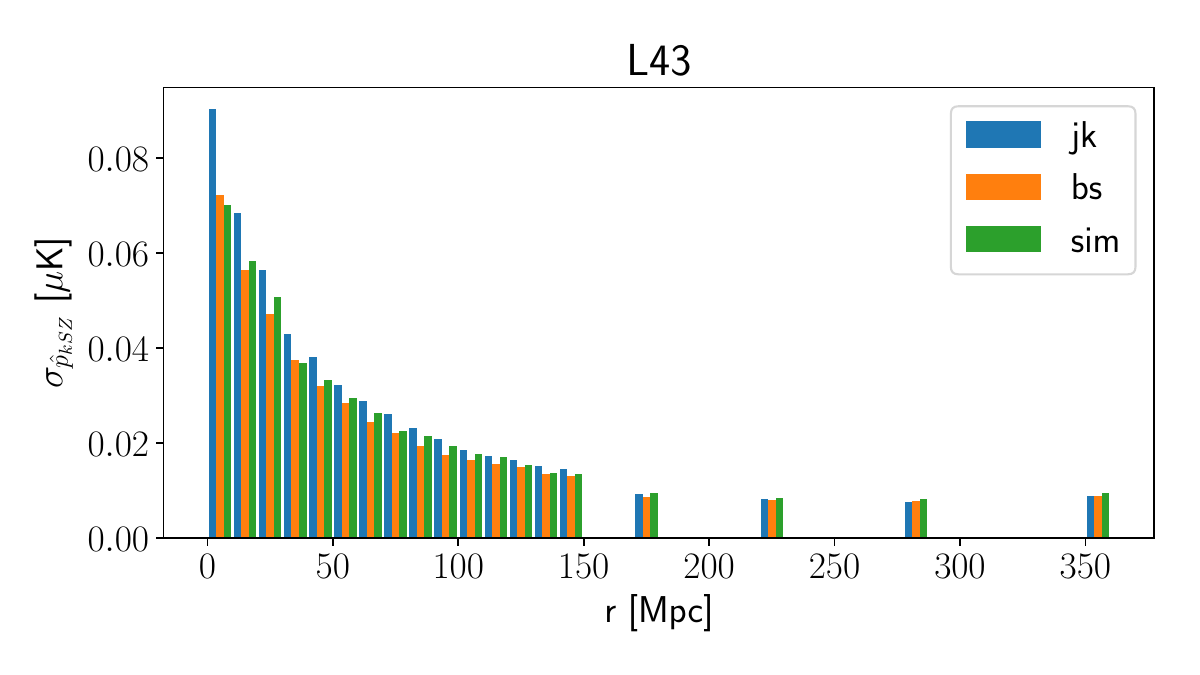}
        	\includegraphics[width=\columnwidth]{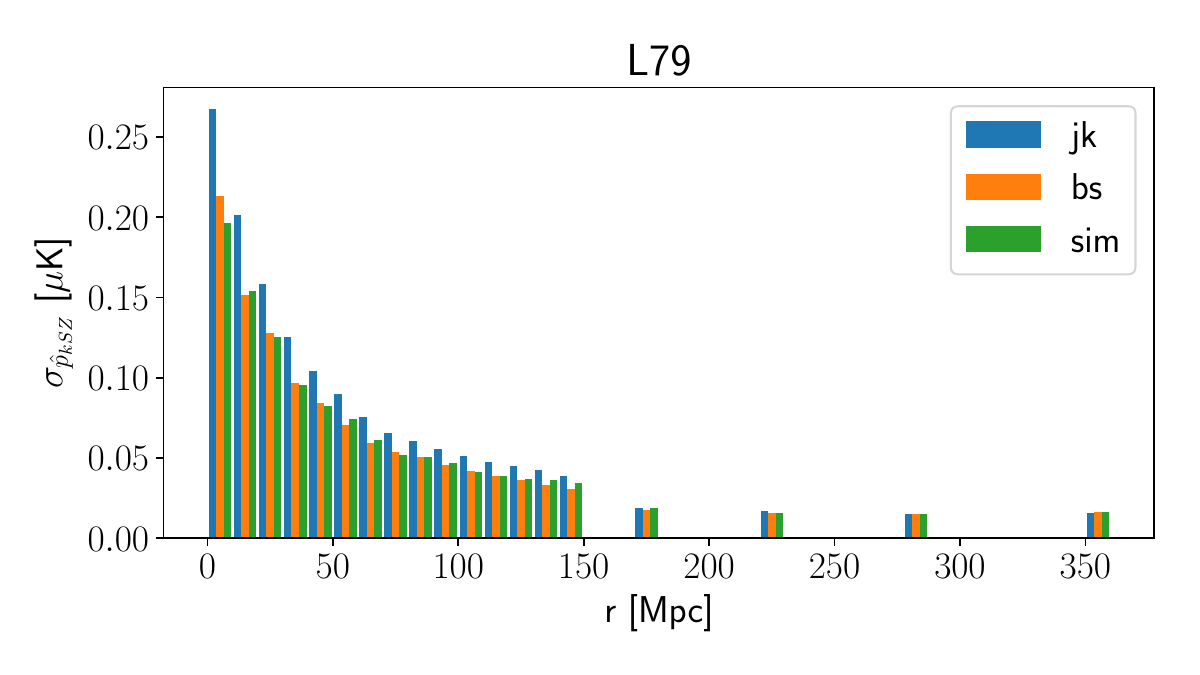}
    \caption{Comparison of the standard deviation of 560 \cs noise simulations (`sim') [green] to the uncertainties obtained from the jackknife (`jk') [blue] and bootstrap (`bs') [orange] resampling of a single noise simulation for [Upper] L43 and [Lower] L79 galaxy tracer samples. }

    \label{fig:jkbs_errorbar_comparison}
\end{figure}

 Figure~\ref{fig:jkbs_errorbar_comparison} shows the standard deviation for the signals obtained from the 560 noise sims, from the JK and bootstrap sampling of the a single noise sim.  The purpose of this comparison is to test the covariance inference technique by checking against simulations. We find that at smaller separations the JK variance exceeds that from noise simulations by 10-25\% with the effect  most pronounced  for the higher luminosity bin, which has comparatively fewer galaxies. We note that the number of galaxy pairs is also comparatively smaller at small versus larger spatial separations.
We conjecture that the trend may be due to the inadequacy of the JK prefactor in (\ref{eq:JK}) to account for the double counting in pair space in the resampling strategy needed for this statistic, when resampling is done in catalog space, especially when fewer pairs are present. We leave further investigation of the origins of these effects, and possible remedies, to a future study. By comparison, the uncertainties obtained from the bootstrap technique  are more consistent with the standard deviations for the noise simulations.
  
\begin{figure}[t!]
	{
	\includegraphics[width=0.50\textwidth]{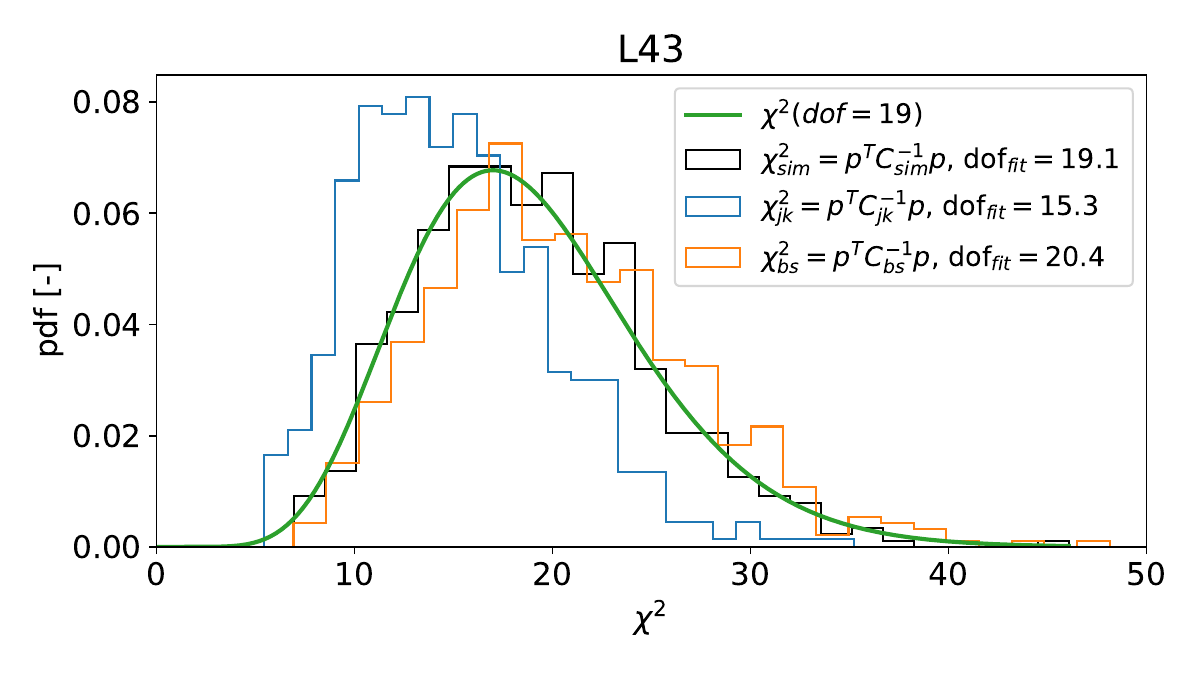}
	\includegraphics[width=0.50\textwidth]{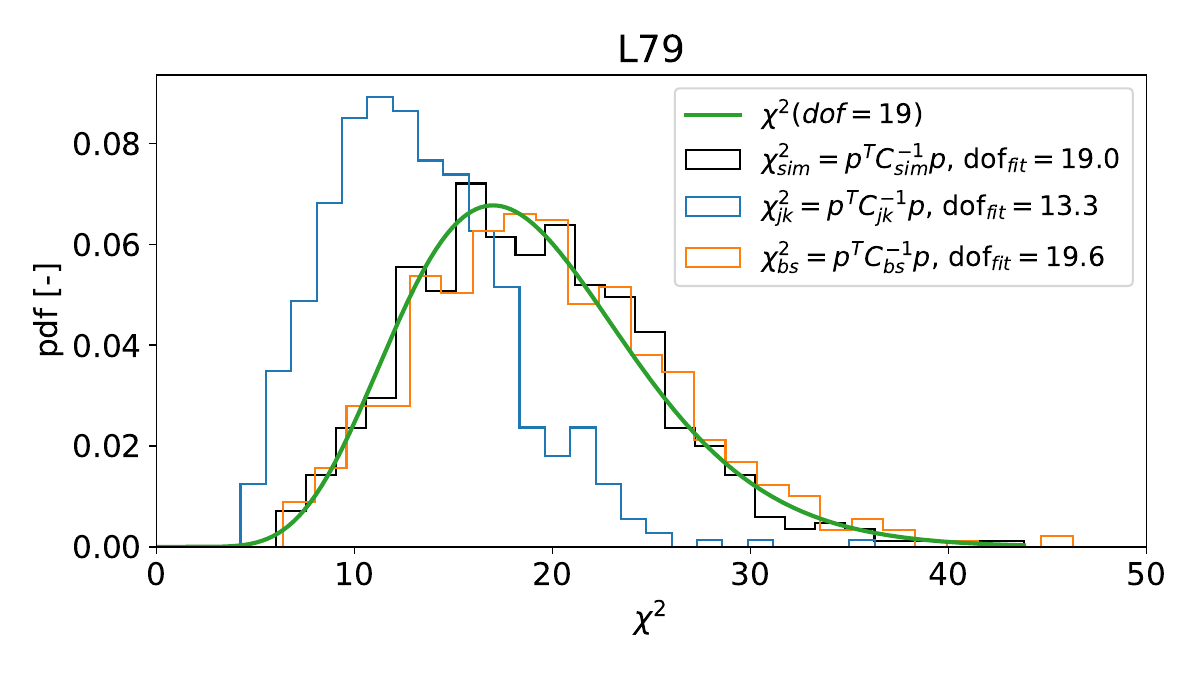}
	}
	\caption{Comparison of the distribution of $\chi^2$ values  obtained from 560 simulation realizations evaluated on covariances inferred using: sims [Black], jackknife (`jk')  [Blue] and bootstrap (`bs') [Orange] resampling of a single sim for the \cs [Upper] L43  and [Lower]  L79 samples versus the theoretical prediction for $\chi^2$ for 19 degrees of freedom (equal to the number of spatial separation bins).}
	\label{fig:chi2comparison}
\end{figure}

  In Fig.~\ref{fig:chi2comparison}, we show the distributions of the $\chi^2$ obtained from the 560 noise sims, and the 1000 resamplings in the JK and bootstrap methods respectively for the 19 spatial separation bins for the L43 and L79 luminosity bins. We find the noise sims well match the expected theoretical $\chi^2$ distribution for 19 degrees of freedom (dof), and that the bootstrap method is comparable, with the best-fit theory having 19.6 and 20.4 dof for  L43 and L79 respectively. The JK method however is found to consistently overestimate the uncertainties leading to the underestimation of the $\chi^2$. The effect is found to be more pronounced in the L79 bin, which has only 30\% as many galaxies as L43; the L79 fit is consistent with a theoretical distribution with 13 dof  while the L43, is slightly better, but still inconsistent, with a theoretical fit of 15 dof.
  
\begin{figure}[t!]
\includegraphics[width=0.5\textwidth]{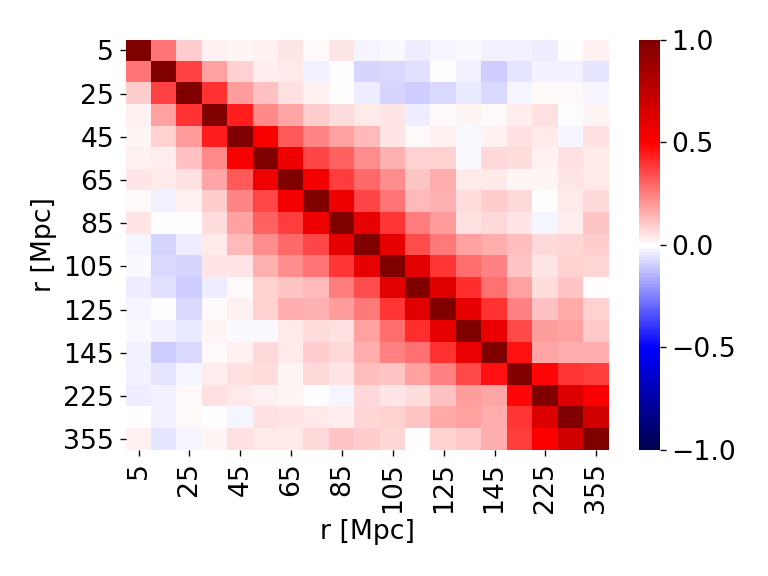}
\includegraphics[width=0.5\textwidth]{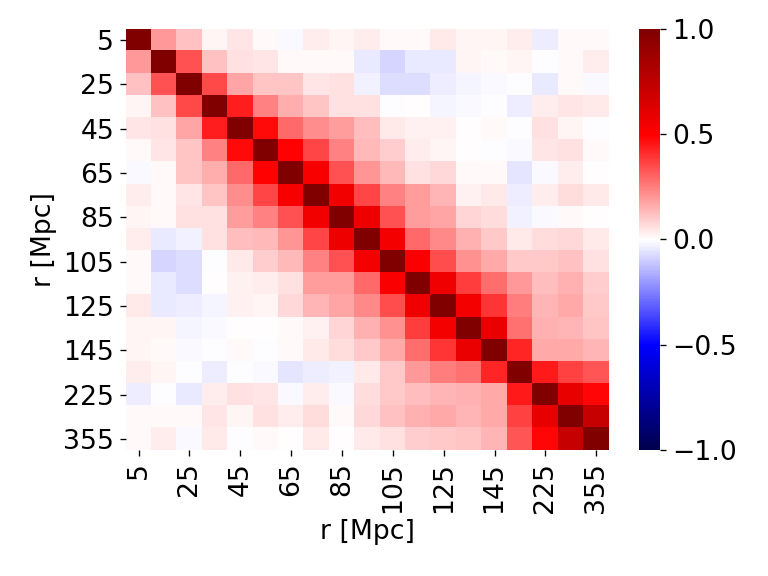}
\includegraphics[width=0.5\textwidth]{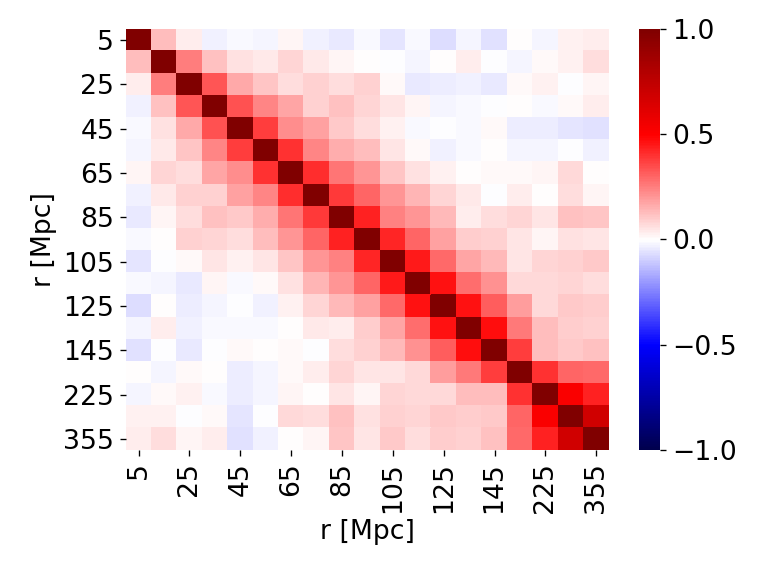}
	\caption{The pairwise correlation matrix, for the \cs map across the 19 spatial cluster separation bins for the  L43  galaxy tracer sample, derived from [top] the covariance across 560 noise sims, [center] the bootstrap and [lower] jackknife resampling.}
	\label{fig:cov}
\end{figure}

In Fig. ~\ref{fig:cov} we show the correlations between the signal across the galaxy separation bins for the L43 sample of the \cs maps estimated  from the bootstrap and JK resampling  of a single noise sim and from the covariance of the 560 noise sims  We find the correlations obtained have similar forms across the three techniques, but that  the bootstrap resampling better captures the correlations beyond adjacent bins found in the correlation matrix from 560 noise sims, while the JK correlation matrix predicts smaller correlations for these pairs.
 
In Table~\ref{tab5} we summarize how the $\bar\tau$ fits vary with the different covariance estimation methods. The PTE values for the JK covariances consistently skew high, especially for the DR5 maps, implying that the JK uncertainties are overestimated. The differences between the JK covariance and that obtained from the 560 sims and bootstrap for the \cs map are most pronounced for the L43D, L61D and L79 samples, in which the sample sizes are the smallest.  The uncertainties in the $\bar\tau$ estimate decrease by 10\% and the best-fit $\bar\tau$ values can shift by up to a half a standard deviation. The larger, cumulative samples, L43 and L61, are  far less impacted by the differences in covariance across the methods, with the best-fit $\bar\tau$ values largely unchanged, and the uncertainties in $\bar\tau$ reduced by $\sim$8\%.  The SNR is also affected with the JK-derived SNR being lower across the board.  The SNR for the \caof map is similar for the L43 and L61 samples, however for the JK covariances the L43 tracer sample has a slightly higher SNR, while for the bootstrap method the best measured signal is for the L61 sample.

\begin{table*}[!t]
  	\begin{tabular}{  | C{4.75em} |  C{5.5em} |C{3.0em} |C{3.0em} |C{3.0em} ||  C{5.5em} |C{3.0em} |C{3.0em} |C{3.0em} || C{5.5em} | C{3.0em} |C{3.0em} |C{3.0em} |}
		 \cline{1-5}
Tracer & \multicolumn{4}{c||}{Simulated noise covariance}&\multicolumn{8}{c}{}
\\ \cline{2-5}
galaxy     &\multicolumn{4}{c||}{\cs }&\multicolumn{8}{c}{}
			 \\ \cline{2-5}   
sample&  $\bar\tau$  ($\times10^{-4}$)&$\chi^2_{\mathrm{min}}$ & PTE  & SNR	&\multicolumn{8}{c}{}
  		\\  \cline{1-5}
$L43D$	& 	0.27 $\pm$ 0.32	&	14	&	0.65	&	0.8	 &\multicolumn{8}{c}{}		
		\\ 
$L61D$  	&  	0.68 $\pm$ 0.32	&	25	&	0.10	&	1.8	&\multicolumn{8}{c}{}
		\\
$L43$ 	& 	0.48 $\pm$ 0.13	& 	20	& 	0.26	&	3.5	&\multicolumn{8}{c}{}
  		\\ 
$L61$ 	& 	0.70 $\pm$ 0.16	&	18	&	0.37	& 	4.1	&\multicolumn{8}{c}{}
  		\\  
 $L79$ 	& 	0.75 $\pm$ 0.23	&	24	&	0.13	&	2.9	&\multicolumn{8}{c}{}

 		\\  \cline{1-5}
\multicolumn{13}{c}{}
\\
\hline	
Tracer & \multicolumn{12}{c|}{ JK covariance}
\\ \cline{2-13}
	 	galaxy     &\multicolumn{4}{c||}{\cs }&\multicolumn{4}{c||}{\can } & \multicolumn{4}{c|}{\caof }
			 \\ \cline{2-13}   
sample&  $\bar\tau$  ($\times10^{-4}$)&$\chi^2_{\mathrm{min}}$ & PTE  & SNR	&  $\bar\tau$  ($\times10^{-4}$)&$\chi^2_{\mathrm{min}}$ & PTE  & SNR		& $\bar\tau$ ($\times10^{-4}$) &$\chi^2_{\mathrm{min}}$ & PTE  & SNR
  		\\ \hline
$L43D$	& 	0.19 $\pm$ 0.35	&	10	&	0.92	&	0.5		&	 0.78 $\pm$ 0.39  	& 7 		& 0.98	& 1.8   	& 	0.53 $\pm$ 0.27 	& 14	 & 0.67 	& 1.7 	
		\\ 
$L61D$  	&  	0.82 $\pm$ 0.37	&	18	&	0.38	&	2.0		&	 1.06 $\pm$  0.40 	& 10 		&  0.92  	& 2.3 	&	 0.71 $\pm$ 0.29 	& 8	& 0.97	& 2.2	
		\\
$L43$ 	& 	0.46 $\pm$ 0.13	& 	17	& 	0.46	&	3.2	 	& 	0.66 $\pm$ 0.15   	& 11		&  0.84	 & 4.2  	& 	  0.54 $\pm$ 0.09 	& 14 	& 0.68 	& 5.1
  		\\ 
$L61$ 	& 	0.77 $\pm$ 0.16	&	14	&	0.64	& 	4.2		& 	0.78 $\pm$  0.18	& 10		& 0.92	& 3.8		& 	0.68 $\pm$ 0.13 	&  9	& 0.95	& 4.8
  		\\  
 $L79$ 	& 	0.67 $\pm$ 0.25	&	16	&	0.53	&	2.3		& 	0.72 $\pm$ 0.31  	& 8		& 0.97  	& 2.1 	& 	0.88 $\pm$ 0.21 	&  9 	& 0.94 	& 3.8	
 		\\ \hline
\multicolumn{13}{c}{}
\\
\hline	
Tracer & \multicolumn{12}{c|}{ Bootstrap covariance}
\\ \cline{2-13}
	 	galaxy     &\multicolumn{4}{c||}{\cs }&\multicolumn{4}{c||}{\can } & \multicolumn{4}{c|}{\caof }
			 \\ \cline{2-13}   
sample&  $\bar\tau$  ($\times10^{-4}$)&$\chi^2_{\mathrm{min}}$ & PTE  & SNR	&  $\bar\tau$  ($\times10^{-4}$)&$\chi^2_{\mathrm{min}}$ & PTE  & SNR		& $\bar\tau$ ($\times10^{-4}$) &$\chi^2_{\mathrm{min}}$ & PTE  & SNR
  		\\ \hline
$L43D$	& 	0.18 $\pm$ 0.32	&	14	&	0.67	&	0.5		&	 0.83 $\pm$ 0.34  	& 12 		& 0.81	& 2.2   	& 	0.46 $\pm$ 0.24 	& 21	 & 0.24 	& 1.7	
		\\ 
$L61D$  	&  	0.69 $\pm$ 0.34	&	25	&	0.08	&	1.8		&	 1.07 $\pm$  0.35 	& 15 		&  0.59  	& 2.7 	&	 0.72 $\pm$ 0.26 	& 11	& 0.85	& 2.5	
		\\
$L43$ 	& 	0.47 $\pm$ 0.12	& 	22	& 	0.20	&	3.6	 	& 	0.65 $\pm$ 0.13   	& 13		&  0.71	 & 4.5  	& 	0.54 $\pm$ 0.09 	& 17 	& 0.42 	& 5.1
  		\\ 
$L61$ 	& 	0.74 $\pm$ 0.15	&	18	&	0.40	& 	4.4		& 	0.82 $\pm$  0.17	& 16		& 0.53	& 4.4		& 	0.69 $\pm$ 0.11 	& 10	& 0.92	& 5.4
  		\\  
 $L79$ 	& 	0.78 $\pm$ 0.23	&	21	&	0.21	&	3.0		& 	0.79 $\pm$ 0.27  	& 12		& 0.79  	& 2.6 	& 	0.88 $\pm$ 0.18 	&  13 & 0.76 	& 4.6	
 		\\ \hline
\multicolumn{13}{c}{}
\end{tabular}  	
		\caption{A comparison of the best-fit $\bar\tau$ estimates and 1$\sigma$ uncertainties  for the \cs [left], \can [center] and \caof [right] maps  for the five luminosity-selected galaxy tracer samples using the noise simulation, bootstrap and jackknife (JK) uncertainty estimates. The upper table shows the \cs results using the covariance obtained across the 560 noise realizations. The center table shows the results using the uncertainties estimated from the JK resampling of the maps, while the lower table shows those using the   bootstrap-derived uncertainties (also presented in Table~\ref{tab2}). The corresponding $\chi^2$ (for 17 degrees of freedom),   SNR and PTE values are also given in each scenario. }
		\label{tab5}
\end{table*}

\bibliographystyle{apsrev}
\bibliography{biblio.bib}
\end{document}